# Inverse Uncertainty Quantification using the Modular Bayesian Approach based on Gaussian Process, Part 1: Theory

Xu Wu[a,*], Tomasz Kozlowski[a], Hadi Meidani[b] and Koroush Shirvan[c]

[a]Department of Nuclear, Plasma and Radiological Engineering, University of Illinois at Urbana-Champaign, Urbana, IL, USA

[b]Department of Civil and Environmental Engineering, University of Illinois at Urbana-Champaign, Urbana, IL, USA

[c]Department of Nuclear Science and Engineering, Massachusetts Institute of Technology, Cambridge, MA, USA

*Phone: (+1) 217-979-7432, Email: xuwu2@illinois.edu

**Abstract**

In nuclear reactor system design and safety analysis, the Best Estimate plus Uncertainty (BEPU) methodology requires that computer model output uncertainties must be quantified in order to prove that the investigated design stays within acceptance criteria. "Expert opinion" and "user self-evaluation" have been widely used to specify computer model input uncertainties in previous uncertainty, sensitivity and validation studies. Inverse Uncertainty Quantification (UQ) is the process to inversely quantify input uncertainties based on experimental data in order to more precisely quantify such ad-hoc specifications of the input uncertainty information.

In this paper, we used Bayesian analysis to establish the inverse UQ formulation, with systematic and rigorously derived metamodels constructed by Gaussian Process (GP). Due to incomplete or inaccurate underlying physics, as well as numerical approximation errors, computer models always have discrepancy/bias in representing the realities, which can cause over-fitting if neglected in the inverse UQ process. The model discrepancy term is accounted for in our formulation through the "model updating equation". We provided a detailed introduction and comparison of the full and modular Bayesian approaches for inverse UQ, as well as pointed out their limitations when extrapolated to the validation/prediction domain. Finally, we proposed an improved modular Bayesian approach that can avoid extrapolating the model discrepancy that is learnt from the inverse UQ domain to the validation/prediction domain.

## 1. Introduction

During the last four decades, the importance of computer simulations has increased dramatically in furthering our understanding of the responses of engineered systems in real world. Large computer codes that implement complex mathematical models have been successfully applied in the design and performance assessment of real systems in many areas of scientific research. Computer modeling is especially significant to the nuclear engineering community, as physical experimentations are usually too costly or sometimes impossible.

### 1.1. Essential components of computer modeling

To bring up the motivation to perform inverse Uncertainty Quantification (UQ), we first briefly establish the definitions of some of the essential components that are used in the credibility evaluation of computer models:

1) *Verification*: "the process of determining that a model implementation accurately represents the developer's conceptual description of the model and the solution to the model" ([1], p. 215). In other words, verification aims to identify, quantify, and reduce errors during the mapping from mathematical model to a computer code.

2) *Code Verification*: the process to access the reliability of the software coding, which includes two activities, *numerical algorithm verification* and *software quality engineering* (SQE) [2]. In other words, code verification deals with adequacy of the numerical algorithms and the fidelity of the computer programming to implement these algorithms.

3) *Solution Verification*: also referred to as *calculation verification* [3], or *numerical error estimation* [2], is the process to evaluate the numerical accuracy of the solutions to a computer code. The primary difference between code and solution verification is that there is generally no known exact solution to the system of interest for the latter. Solution verification strongly depends on the quality and completeness of code verification, and both processes should be performed prior to validation, as defined below.

4) *Validation*: "the process of determining the degree to which a model is an accurate representation of the real world from the perspective of the intended uses of the model" ([1], p. 215). In other words, validation aims to determine the degree of accuracy of the considered model in representing real world phenomena. Verification and Validation together are often termed "V&V".

5) *Forward UQ*: the process of quantifying the uncertainties in Quantity-of-Interest (QoIs) by propagating the uncertainties in input parameters through the computer model [4][5]. QoIs predictions along with uncertainties are necessary for validation.

6) *Sensitivity analysis (SA)*: the study of how uncertainties in the QoIs of can be apportioned to various random input parameters [6]. SA provides a ranking of the input parameters by their significance to QoIs.

7) *Optimization*: the process of maximizing or minimizing an object function by systematically choosing input values from within an allowed set [7][8].

8) *Calibration*: the process of adjusting a set of input parameters implemented in the code so that the agreement of the computer code predictions with corresponding experimental data is maximized [3].

9) *Data Assimilation*: the process to incorporate observations of the actual system into the model state of a numerical model of that system [9]. Data assimilation can be treated as the calibration of dynamic models, which arise in many fields of geosciences such as weather forecasting.

10) *Benchmark*: "A benchmark is a choice of information that is believed to be accurate or true for use in verification, validation or calibration" ([3], p. 1333). For example, benchmarks can be measurements of QoIs from physical experiments or solutions from highly accurate numerical tests.

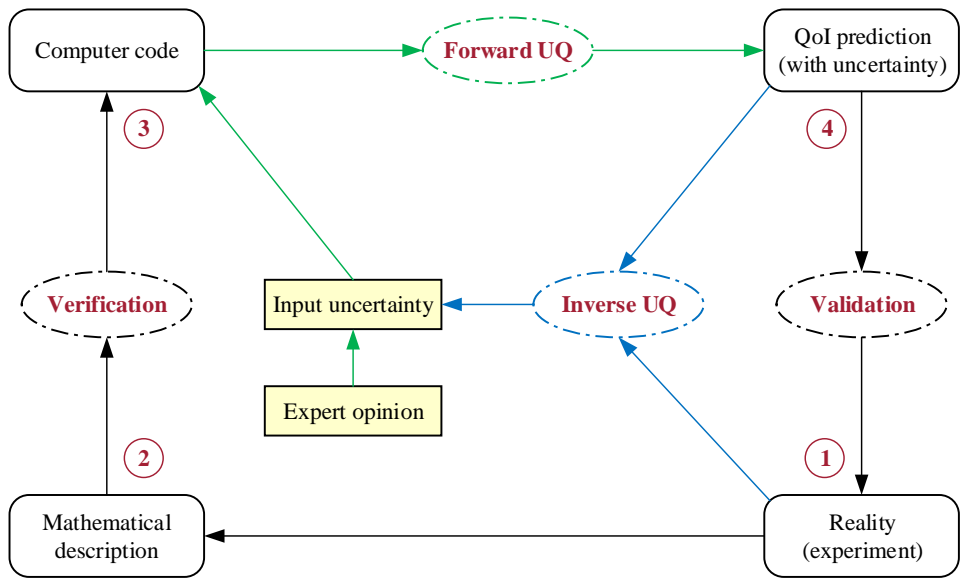

Figure 1: Some essential parts of computer modeling (a non-exclusive list)

Figure 1 shows the connections between some of these essential components of computer modeling. From Figure 1 it is obvious that the forward UQ process always starts with characterization of the input uncertainties, for example, the mean values, variances, Probability Density Functions (PDFs), upper and lower limits, etc. Unfortunately, such information is not always readily available to the code users. Such condition is known as the "lack of input uncertainty information" issue. Up to now, in the uncertainty, sensitivity and validation studies of nuclear engineering, "expert opinion" or "user self-assessment" have been predominantly used (see reviews in [10][11]). Such ad-hoc specifications of input uncertainty information have been considered reasonable for a long time. However, these approaches are subjective and lack mathematical rigor, and can lead to inconsistencies.

The "lack of input uncertainty information" issue necessitates the research on inverse UQ. An early appearance of the term "inverse UQ" can be found in [1], in which it was also termed "backward problem". Other researchers have called it "inverse uncertainty propagation" [12]. According to Oberkampf and Trucano, "The backward problem asks whether we can reduce the output uncertainty by updating the statistical model using comparisons between computations and experiments" ([1], p. 256). In this paper, we will introduce the theory for inverse UQ under the Bayesian framework in an evolving manner, including the Bayesian formulation for inverse UQ, Gaussian Process (GP) metamodeling, full and modular Bayesian approaches, and finally an improved modular Bayesian approach.

**1.2. Inverse UQ vs. calibration**

Inverse UQ, also referred to as *inverse problem* or *parameter estimation*, is the process to quantify the uncertainties of input parameters based on chosen experimental data. Such definition looks very similar with calibration. In this subsection we briefly discuss the relationship between inverse UQ and calibration.

Calibration can be classified as deterministic and statistical calibration [13]. *Deterministic calibration* merely determines the point estimates of best-fit input parameters such that the discrepancies between code output and experimental data can be minimized. However, *statistical calibration*, sometimes referred to as *Bayesian calibration* [14], *probabilistic inversion* [15] or *Calibration under Uncertainty (CUU)* [3], produces statistical descriptions like distributions. In this sense, inverse UQ is same with Bayesian calibration and indeed they do share the same techniques. For example, both of them employ the Bayesian inference theory [16] and explore the posterior PDF with Markov Chain Monte Carlo (MCMC) sampling [17]. They both favor surrogate models when the computational models are expensive. So what makes inverse UQ in the current study different from Bayesian calibration?

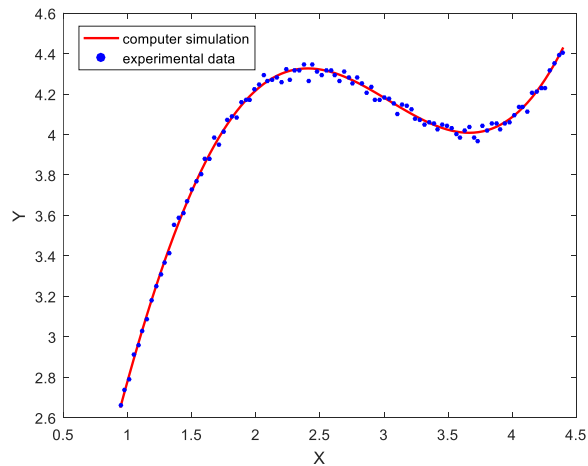

Figure 2: A simple case when calibration will not improve the agreement between simulation and measurement.

Inverse UQ only has very subtle differences with Bayesian calibration, (1): inverse UQ includes some techniques that implements the Expectation-Maximization (E-M) algorithm [18] rather than sampling of the posterior PDF, even though the former is not as widely applicable as the latter; (2): they are usually performed with different motivations. Bayesian calibration aims at reducing the difference between simulation and observation, while inverse UQ emphasizes quantifying the input uncertainties. When the model outputs already agree very well with experimental data, we may conclude that no calibration is needed. However, the inverse UQ is still useful because the underlying uncertainties in model input parameters have to be quantified. Figure 2 illustrates such a case, when the differences between simulation and measurement approximately follow Gaussian noise with a very small variance. In that case, calibration is unlikely to improve the agreement between simulation and observation. In essence, in cases where there is no need to do Bayesian calibration, inverse UQ may still be useful.

The advantage of inverse UQ (or Bayesian calibration) over deterministic calibration and "parameter tuning" is apparent: (1): firstly, information on QoIs from experiments is never sufficiently accurate to allow inference of the "true" or "exact" values of the input parameters. Instead, we can only hope to reduce our ignorance of the parameters by achieving less uncertainties in them (the so-called uncertainty reduction); (2): furthermore, it is difficult for deterministic calibration to quantify correlations among the estimates; (3): thirdly, it is highly possible that various combinations of input parameters will yield simulations that have similar agreement with the measurement data. Deterministic calibration, which relies on optimization techniques to select best-fit values, may end up with getting only one of a set of equally well-fitting values. This is especially true for over-parameterized models given limited observations [15]; (4): finally, the observed data usually contains certain degree of uncertainty, which should be considered during the inference process of calibration parameters.

Inverse UQ problems have received increasing attention in the modeling & simulation community, primarily in the context of source inversion and calibration of model input parameters. Some representative examples are identification of source term and deposition velocity in nuclear radiation release [14], source inversion in transient diffusion [19][20], source inversion in heat conduction and permeability estimation in flow through porous media [21], recovery of the location and intensity of a radiation source in the urban area [22]. Other applications can be found for simply supported beam [23], charged particle accelerator and spot welding experiment [24], shock physics, materials science, cosmology, and particle physics [25], thermally decomposing foam [26], process-based forest models [15] and climate model with terrestrial carbon cycle [27], etc.

### 1.3. Motivation for forward/inverse UQ in nuclear engineering

Historically in nuclear system design and safety assessment, computer codes used extreme or unfavorable values of input parameters to produce conservative predictions of the responses. Such an approach quantified reactor designs with a considerable margin to assure its safety and avoid under-prediction of safety-related outputs (e.g. peak cladding temperature (PCT)). It did so by modeling the physical phenomena at the worst-case scenario. Consequently, the conservative approach typically lead to considerable inaccuracy in simulation and ultimately damaged the economic performance of nuclear energy.

In the 1980s, best estimate safety analysis strategy started to be embedded in the Code Scaling Applicability and Uncertainty (CSAU), and Phenomena Identification and Ranking Table (PIRT) methodologies, which were accepted by the U.S. Nuclear Regulatory Commission (USNRC). This strategy is commonly referred to as the Best Estimate plus Uncertainty (BEPU) methodology [28]. The goal of BEPU methodologies aims to capture the physical phenomena as realistically as possible by implementing a wide range of modeling options and accurate calculation methods to capture physical phenomena at a greater fidelity. According to the BEPU methodology, uncertainties must be quantified in order to prove that the investigated design stays within acceptance criteria.

Validation, forward and inverse UQ play a more significant role in nuclear engineering compared to other fields to reduce conservatism while dealing with high-consequence systems. The design decision-making process, development of public policy and preparation of safety procedures all rely on reliable computer codes that have undergone extensive V&V process and proven to be of high credibility. We will provide a detailed review of representative applications of calibration and inverse UQ in nuclear engineering in a companion paper [29]. Given the limited work on inverse UQ among the nuclear community, we devote this paper and the companion paper to this topic, hoping to draw more attention into this area and raise the interest on these topics to a higher level in our community.

### 1.4. Objective and outline

This paper and the companion paper [29] describe the process of performing inverse UQ using the modular Bayesian approach. We aim to provide sufficient details so that the readers can readily implement the methodology without consulting other materials. However, pointers to the related literature will also be included. This paper will focus on introducing the theory for inverse UQ using the modular Bayesian approach. Topics concerning the Bayesian formulation of inverse UQ, metamodel construction and validation using GP, full, modular and an improved modular Bayesian approach will be covered. The companion paper [29] will demonstrate the application of the improved modular Bayesian approach outlined in this paper to nuclear reactor system thermal-hydraulics code TRACE. In that paper topics including test source allocation (TSA) and mathematical description of model discrepancy term will also be covered. Note that within the scope of this paper and its companion paper, we will assume that verification has been carried out on the computer codes to produce converged solutions throughout the domain of application.

This paper is organized in the following way. Section 2 establishes the Bayesian formulation for inverse UQ. In Section 3, the theory for GP modeling will be presented. Section 4 discusses full and modular Bayesian approach, as well as their comparison and limitations. Finally, a lightweight and easy-to-use version of the modular Bayesian approach will be introduced in Section 5. Section 6 concludes this paper.

## 2. Formulation of Inverse UQ Problem

In this section, the formulation of the inverse UQ problem for a general computer model is provided. We start with a classification of inputs parameters. Then the "model updating equation" is introduced that incorporates the model discrepancy term. Finally, the Bayesian solution for inverse UQ is presented.

### 2.1. Classification of input parameters

Consider a general computer model $y^M = y^M(\mathbf{x}, \boldsymbol{\theta})$ where $y^M$ is the model output (also called response or QoI) which can be either a scalar or vector that corresponds to multi-dimensional outputs. The vector $\mathbf{x} = [x_1, x_2, ..., x_r]^T$ is the vector of *design variables* (also called *system inputs*, *control variables*, or *observable variables*), and $\boldsymbol{\theta} = [\theta_1, \theta_2, ..., \theta_d]^T$ is the vector of *calibration parameters* (sometimes called *ancillary variables*). Examples of design variables are initial conditions (ICs) and boundary conditions (BCs). For instance, in our application to TRACE [29], the design variables are pressure, mass flow rate, power and inlet temperature. Calibration parameters are specified as inputs to the computer model but are unknown or not measurable when conducting the physical experiments. In this work we use a broad definition of calibration parameters similar to [27], which includes:

1) Physical constants, such as physical model parameters like material properties and heat transfer coefficients (HTC);

2) Tuning parameters, which are needed to make the model perform well, like multiplicative or additive factors. Tuning parameters are usually notional and of little or no physically interpretable meaning.

3) Context-specific constants, such as switch between different scenarios. For example, switch between various flow regimes in nuclear reactor system thermal-hydraulics analysis.

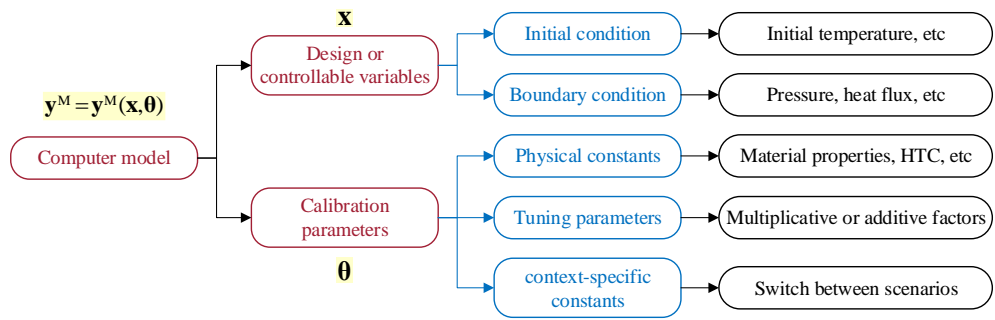

Figure 3: Classification of input parameters for a general computer model.

Figure 3 shows the classification of input parameters for a general computer model. We would like to point out a few distinctions between **x** and **θ** as it is important to avoid the confusion before inverse UQ:

1) Design variables are usually required by both computer simulation and physical experimentation, while calibration parameters are only needed by the former.

2) Design variables usually have clear and unambiguous physical meaning, while calibration parameters may have a physical meaning in nature or be purely numerical.

3) Design variables are used to describe the conditions or scenarios under which the experiments have been performed, while the calibration parameters have inherent values that remain unchanged under different scenarios or experimental conditions [13].

4) Design variables are usually assumed to be known or at least observable during experimentation. These variables may also be subject to uncertainties due to known "variability" that may be reported along with the benchmark. Calibration parameters, however, have unknown uncertainties that are usually characterized by prior or posterior distributions representing epistemic uncertainty rather than variability.

5) The distinction between **x** and **θ** is not important for many purposes like forward UQ and sensitivity analysis. But in inverse UQ, calibration parameters are the quantities that are estimated.

It has to be noted that there are some different classifications on model inputs in previous work on calibration, validation or model updating. In the work of Campbell [13], "inputs" and "parameters" were used to represent design variables and calibration parameters respectively. However, this can easily cause confusion for the readers. Therefore, we assume no difference between "input", "variable" and "parameter" and use "design" and "calibration" in front of these terms to explicitly refer to **x** and **θ**. In the work of Kennedy and O'Hagan [14], design variables were also called "variable inputs". Finally, some researchers [30] treated tuning and calibration as separate processes. In this work, like many other previous research [24][31][32][33], we make no such distinction between tuning and calibration parameters and present an approach that results in post-calibration distributions for both types.

## 2.2. Model updating equation

We use "experiment", "observation" and "measurement" interchangeably in this work, assuming no difference between them and also use the terms "physical" or "field" in front of them in order to be consistent with the open literature. The connections between computer model simulation, reality and experiments are illustrated in Figure 4. Given an experimental condition characterized by design variables **x**, to learn about the real or true value of the QoIs $y^R(\mathbf{x})$, we run the computer model to obtain the model prediction $y^M(\mathbf{x}, \boldsymbol{\theta}^*)$. This model prediction is obtained using the parameter value $\boldsymbol{\theta}^*$, referred to as the "best[1]" or "true" but unknown values for the calibration parameters. The main goal of inverse UQ process is to determine this "best" value $\boldsymbol{\theta}^*$.

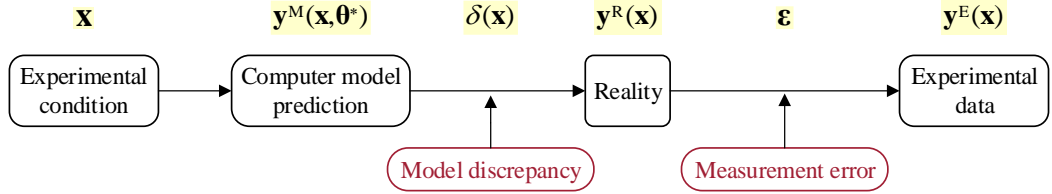

Figure 4: The connections between computer model prediction, reality and experimental data.

Because computer models can inevitably provide only approximations of the reality, the term $\delta(\mathbf{x})$ is introduced to represent the discrepancy between computer simulation and reality. $\delta(\mathbf{x})$ is called the *model uncertainty*, also called *model discrepancy*, *model inadequacy* or *model bias/error* [14][32]. $\delta(\mathbf{x})$ is due to incomplete or inaccurate underlying physics, numerical approximation errors, and/or other inaccuracies that would exist even if all the parameters in the computer model were accurately and deterministically specified. We now have the following relation:

$$y^R(\mathbf{x}) = y^M(\mathbf{x}, \boldsymbol{\theta}^*) + \delta(\mathbf{x}) \qquad (1)$$

To directly measure the reality $y^R(\mathbf{x})$, we also perform experiments to obtain observation $y^E(\mathbf{x})$. In the experimental process we will inevitably have (typically additive) measurement error/noise:

$$y^E(\mathbf{x}) = y^R(\mathbf{x}) + \boldsymbol{\varepsilon} \qquad (2)$$

where $\boldsymbol{\varepsilon} \sim \mathcal{N}(\boldsymbol{\mu}, \boldsymbol{\Sigma}_{\exp})$ represents an example of an additive measurement error, here taken to be normally distributed.. Note that there can be multiple measurements and it is widely accepted to have homoscedastic experimental errors $\boldsymbol{\Sigma}_{\exp} = \sigma^2_{\exp}\boldsymbol{I}$. Also, $\boldsymbol{\mu} = \boldsymbol{0}$ is frequently used, assuming that the instrumentation has no systematic bias and the mean value of the measurement is the same as the real response.

---

[1] By "best" we mean that the model run at $\boldsymbol{\theta}^*$ gives the most accurate prediction. However, it has to be noted that the "best" value may be different from the "real" value. $\boldsymbol{\theta}^*$ is the "best" value only in the sense of most accurately representing the measurement data. Due to model discrepancy, model prediction may not agree well with reality when the model runs at the "real" value. However, since the "real" value can never be learnt, by convention the "best" value and "real" value are treated as the same.

The model discrepancy term was first addressed in the seminal work of Kennedy and O'Hagan [14]. It is important to consider model discrepancy $\delta(\mathbf{x})$ as otherwise we would have an unrealistic level of confidence in the computer simulations [32]. Equation (1) shows that without $\delta(\mathbf{x})$ we will have "model = reality", which is not reasonable and will cause "*over-fitting*". Over-fitting means that the calibration parameters closely match certain set of experiments to a point that the computer code may perform poorly when applied to other experiments. In the companion paper [29] we will demonstrate that the introduction of the model discrepancy term into the methodology presented in this paper can in fact help avoid over-fitting. By combining Equations (1) and (2) we have:

$$\mathbf{y}^{\mathrm{E}}(\mathbf{x}) = \mathbf{y}^{\mathrm{M}}(\mathbf{x}, \boldsymbol{\theta}^*) + \delta(\mathbf{x}) + \boldsymbol{\varepsilon} \tag{3}$$

Equation (3) is frequently referred to as "*model updating formulation/equation*" [23]. The model updating equation will serve as the starting point of inverse UQ. In the next section, we will use Bayesian inference and Equation (3) to estimate the probability distribution for the "true" values $\boldsymbol{\theta}^*$ of the calibration parameters, using observation data.

### 2.3. Bayesian solution for inverse UQ problem

The Bayesian inference theory [16] is used to determine the posterior PDF of the "true" calibration parameters $\boldsymbol{\theta}^*$ which is defined as $\mathrm{p}(\boldsymbol{\theta}^*|\mathbf{y}^{\mathrm{E}}, \mathbf{y}^{\mathrm{M}})$, where $\mathbf{y}^{\mathrm{E}}$ and $\mathbf{y}^{\mathrm{M}}$ are ensembles of field observations and computer simulations respectively. According to the Bayesian theory we have:

$$\mathrm{p}(\boldsymbol{\theta}^*|\mathbf{y}^{\mathrm{E}}, \mathbf{y}^{\mathrm{M}}) \propto \mathrm{p}(\mathbf{y}^{\mathrm{E}}, \mathbf{y}^{\mathrm{M}}|\boldsymbol{\theta}^*) \cdot \mathrm{p}(\boldsymbol{\theta}^*) \tag{4}$$

where $\mathrm{p}(\boldsymbol{\theta}^*)$ is the prior PDF of the parameter, and $\mathrm{p}(\mathbf{y}^{\mathrm{E}}, \mathbf{y}^{\mathrm{M}}|\boldsymbol{\theta}^*)$ is the likelihood function. In brief, prior and posterior PDFs, or in short prior and posterior, represent degrees of belief about possible values of $\boldsymbol{\theta}^*$, before and after observing the experimental data $\mathbf{y}^{\mathrm{E}}$. Given a particular value for $\boldsymbol{\theta}^*$, the likelihood function measures the probability of the observed data $\mathbf{y}^{\mathrm{E}}$ being associated with it. Once we have a specification for the prior and likelihood function, one can identify the posterior distribution of parameters, not by directly solving for the posterior PDF, but by obtaining samples that are distributed according to the posterior. This is predominantly done using Markov chain Monte Carlo (MCMC) [17]. The MCMC samples can be then used to build the posterior PDF, if they are desired. Analytical (direct) solutions of the posterior may also be possible in rare conditions when the prior and likelihood are sufficiently simple.

When model discrepancy is also considered, the formulation of the likelihood function $\mathrm{p}(\mathbf{y}^{\mathrm{E}}, \mathbf{y}^{\mathrm{M}}|\boldsymbol{\theta}^*)$ is a challenging task, as we need proper definitions for model discrepancy. We have previously treated the measurement error $\boldsymbol{\varepsilon}$ as independent and identically distributed (i.i.d.) zero-mean Gaussian noise, whose variance is expected to be reported along with measurement data. This is easily done since the error rates for most instruments are known and accessible. The model discrepancy $\delta(\mathbf{x})$, on the other hand, needs a proper formulation. This is still an area of active research [34]. GP with uninformative priors on the parameters in $\delta(\mathbf{x})$ is a popular choice [14][24][25][27]. However, such formulation is usually problem specific [27]. A more detailed and complete discussion of the model discrepancy term will be presented in Section 4 where we introduce the modular Bayesian approach.

Besides a proper formulation, another important issue associated with model discrepancy $\delta(\mathbf{x})$ is called the "*identifiability*" [23]. Identifiability concerns the question whether the "true" value $\boldsymbol{\theta}^*$ can theoretically be inferred based on the available measurement data. Identifiability is not trivial as it is difficult to know how much of the difference between computer simulations and field experiments should be attributed to the uncertainty in $\boldsymbol{\theta}^*$, model discrepancy $\delta(\mathbf{x})$ or measurement error $\boldsymbol{\varepsilon}$. In other words, multiple combinations of uncertainties due to $\boldsymbol{\theta}^*$, $\delta(\mathbf{x})$ and $\boldsymbol{\varepsilon}$ can equally well explain the same mismatch between model prediction and field measurements, making the "true" value $\boldsymbol{\theta}^*$ not (uniquely) identifiable.

Given the above discussion, $\boldsymbol{\varepsilon} = \mathbf{y}^{\mathrm{E}}(\mathbf{x}) - \mathbf{y}^{\mathrm{M}}(\mathbf{x}, \boldsymbol{\theta}^*) - \delta(\mathbf{x})$ follows a multi-dimensional Gaussian distribution. The posterior can be written as:

$$\mathrm{p}(\boldsymbol{\theta}^*|\mathbf{y}^{\mathrm{E}}, \mathbf{y}^{\mathrm{M}}) \propto \mathrm{p}(\boldsymbol{\theta}^*) \cdot \frac{1}{\sqrt{|\boldsymbol{\Sigma}|}} \exp\left[-\frac{1}{2}[\mathbf{y}^{\mathrm{E}} - \mathbf{y}^{\mathrm{M}} - \delta]^{\mathrm{T}} \boldsymbol{\Sigma}^{-1} [\mathbf{y}^{\mathrm{E}} - \mathbf{y}^{\mathrm{M}} - \delta]\right] \tag{5}$$

If heteroscedastic experimental errors are assumed between different measurements or different QoIs, we should use $\boldsymbol{\varepsilon} \sim \mathcal{N}(\mathbf{0}, \boldsymbol{\Sigma}_{\mathrm{exp}})$. The diagonal entries represent the variances for each error component while the off-diagonal elements are their covariance. Note that $\mathbf{\Sigma}_{\text{exp}}$ is only part of the likelihood covariance matrix $\mathbf{\Sigma}$, which includes other sources of uncertainties as shown in Figure 5. We have:

$$\mathbf{\Sigma} = \mathbf{\Sigma}_{\text{exp}} + \mathbf{\Sigma}_{\text{bias}} + \mathbf{\Sigma}_{\text{code}} \qquad (6)$$

where $\mathbf{\Sigma}_{\text{exp}}$ is the *experimental uncertainty* caused by measurement noise. The second term $\mathbf{\Sigma}_{\text{bias}}$ represents the *model uncertainty* due to, as we have discussed in Section 2.2, incomplete/inaccurate underlying physics and numerical approximation errors. The third term $\mathbf{\Sigma}_{\text{code}}$ is called *code uncertainty*, or *interpolation uncertainty*, because we do not know the computer code outputs at every input, especially when the code is computationally prohibitive. In this case, we might choose to use some kind of metamodels. In Section 3.3 we will see that GP is a very good choice for metamodels as it provides an estimation of the code uncertainty $\mathbf{\Sigma}_{\text{code}}$.

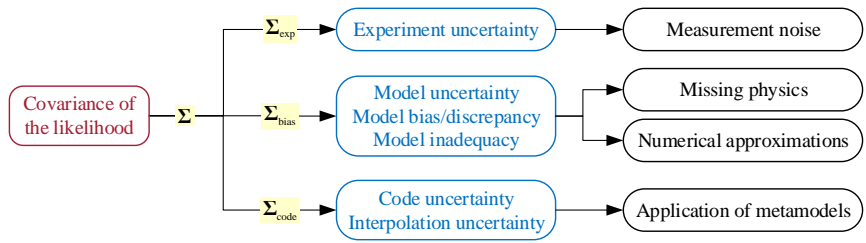

Figure 5: Components of the covariance matrix of the likelihood function.

## 2.4. Motivation for using metamodels

The posterior PDF $p(\boldsymbol{\theta}^*|\boldsymbol{y}^E, \boldsymbol{y}^M)$ is the Bayesian solution to the inverse UQ problem. Posterior PDF represents the uncertainty about $\boldsymbol{\theta}^*$ informed by available experimental data. Various statistical moments and probability densities can be computed as long as we manage to generate samples that are distributed according to the posterior PDF $p(\boldsymbol{\theta}^*|\boldsymbol{y}^E, \boldsymbol{y}^M)$. MCMC sampling is commonly used to draw posterior samples. The main advantage of MCMC is its ability to generate samples from a PDF that is known up to a normalizing constant. The most important issue with using MCMC together with expensive codes, is that MCMC requires sufficiently large number of samples to fully explore the posterior PDF. This means numerous full model execution, i.e. calls to computationally prohibitive codes, are required by MCMC. This issue is frequently bypassed by using metamodels.

Metamodels are approximations of the input/output relations of computer models. They are also called *surrogate models*, *response surfaces* or *emulators*[2]. They are built from a limited number of runs of the full simulation code at specially selected values of the random input parameters (the so-called experimental design [35][36]) and a learning algorithm. Metamodels usually take much less computational time than the full model while maintaining the input/output relation to a desirable accuracy. Accurate metamodels can be used to replace the full models in uncertainty, sensitivity, validation and optimization studies, etc.

Typical examples of metamodels include Moving Least-Squares (MLS) [37], Radial Basis Functions (RBF) [38], Artificial Neural Network (ANN) [39], Support Vector Machine (SVM) [40], Gaussian Process (GP) [35][36], generalized Polynomial Chaos Expansion (PCE) [41], Sparse Grid Stochastic Collocation (SGSC) [42], etc. See [7][8] for detailed reviews of metamodels. Recently, metamodels that are built with stochastic spectral methods like PCE and SGSC [10][11][19][20][21] and GP emulators [14][15][24][25][26][27][43][44][45] have been increasingly used in facilitating Bayesian inference/calibration when expensive model executions are involved.

## 3. Gaussian Process Metamodeling

GP modeling, also known as *Kriging*, or *spatial correlation modeling* [46], was originally developed by geologists in the 1950s to predict distribution of minerals over an area of interest given a set of sampled sites [47][48][49]. It was made popular in the context of modeling and optimization by Sacks et al. [35][50] and Jones et al. [51]. GP has been widely used to construct metamodels for deterministic computer models in many areas [46][52]. A GP model is a generalized linear regression model that accounts for the correlation in the residuals between the regression model and the observations. In the case of metamodeling, the "observations" are simulation outputs of computer codes at selected values of the inputs, rather than experimental data. The only assumption for GP modeling is that the model output is continuous and smooth over the input domain [7], which is true for most computer models. We thus believe that if two input points are close to each other in the input domain, the residuals in the regression model should be close [51]. It follows that we do not treat the residuals as independent, but assume that the correlation between the residuals are related to the distance between the corresponding input points.

---
[2] The term "emulator" (as a comparison, the original or full model is called a "simulator") is often used for probabilistic response surfaces whose estimation at an untried input is a distribution rather than a point value. Therefore an emulator is a statistical approximation of the simulator.

In this paper, we will present a self-contained introduction of GP, including the general theory, correlation kernels, prediction and Mean Square Error (MSE) formula, design of computer experiments, parameter estimation and accuracy assessment. However, we admit that the richness of the GP modeling filed is just hinted in this section, given the fact that several books have been devoted on this topic [36][45][47][48][49].

### 3.1. General theory of GP

Consider a mathematical model of the general form $y = y^M(\mathbf{x})$ (not to be confused with the definition of computer model in Section 2.1, here $\mathbf{x}$ can include both design and calibration parameters). Without loss of generality, assume that y is a scalar and $\mathbf{x} = [x_1, x_2, \ldots, x_d]$ representing $d$ input parameters. Assume that the computer model output is known at $m$ design sites (also called training sites):

$$\mathbf{X} = \left[\mathbf{x}^{(1)}, \mathbf{x}^{(2)}, \ldots, \mathbf{x}^{(m)}\right]^T \tag{7}$$

where $\mathbf{X}$ is a $m \times d$ design matrix. The corresponding output values are:

$$\mathbf{y} = [y_1, y_2, \ldots, y_m]^T = \left[y^M(\mathbf{x}^{(1)}), y^M(\mathbf{x}^{(2)}), \ldots, y^M(\mathbf{x}^{(m)})\right]^T \tag{8}$$

The mathematical form of a GP model is given by:

$$y(\mathbf{x}) = \sum_{j=1}^{n} \beta_j f_j(\mathbf{x}) + z(\mathbf{x}) = \mathbf{f}^T(\mathbf{x})\boldsymbol{\beta} + z(\mathbf{x}) \tag{9}$$

Here $y(\mathbf{x})$ is the output prediction at a general location denoted by $\mathbf{x}$. The first part is a linear regression of the data with $n$ repressors modeling the drift of the mean, also called the *"trend"*. The set of basis functions $\mathbf{f} = [f_1, f_2, \ldots, f_n]^T$ are known and chosen by the user, while the vector $\boldsymbol{\beta} = [\beta_1, \beta_2, \ldots, \beta_n]^T$ contains the regression coefficients. The second part $z(\mathbf{x})$ is a stationary Gaussian random process with zero mean and covariance:

$$\text{Cov}\left[z(\mathbf{x}^{(i)}), z(\mathbf{x}^{(j)})\right] = \sigma^2 \mathcal{R}(\mathbf{x}^{(i)}, \mathbf{x}^{(j)}) \tag{10}$$

where $\sigma^2$ is a scalar parameter called *process variance* and $\mathcal{R}(\cdot,\cdot)$ is called the *(spatial) correlation function* or *correlation kernel*, defined for any two points, $\mathbf{x}^{(i)}, \mathbf{x}^{(j)}$, in the input domain. The inclusion of the second part $z(\mathbf{x})$ generalizes the linear regression model to a "*stochastic process model*".

### 3.2. Correlation kernels

The correlation function/kernel is often chosen to be a function of the distance:

$$\mathcal{R}(\mathbf{x}^{(i)}, \mathbf{x}^{(j)}) = \mathcal{R}[d(\mathbf{x}^{(i)}, \mathbf{x}^{(j)})] \tag{11}$$

Instead of using the Euclidean distance that weights all the variables equally, we use a specially weighted distance formula defined below:

$$d(\mathbf{x}^{(i)}, \mathbf{x}^{(j)}) = \sum_{k=1}^{d} \frac{|x_k^{(i)} - x_k^{(j)}|^{p_k}}{\omega_k} \tag{12}$$

For example, if we choose the power-exponential function for the correlation kernel $\mathcal{R}(\cdot,\cdot)$, the covariance becomes:

$$\text{Cov}[z(\mathbf{x}^{(i)}), z(\mathbf{x}^{(j)})] = \sigma^2 \cdot \exp\left[-\sum_{k=1}^{d} \frac{|x_k^{(i)} - x_k^{(j)}|^{p_k}}{\omega_k}\right] \tag{13}$$

The parameters $\mathbf{p} = [p_1, p_2, \ldots, p_d]$ are called *roughness parameters*, which control the smoothness of the correlation function. It is usually required that $p_k \in [1, 2]$ [51], but a wider range $p_k \in [0, 2]$ has also been used [7][45][53]. It can be seen from Figure 6 (left) that when the distance $h$ between two points approaches 0, the correlation gets closer to 1. And the correlation decreases when the distance gets larger. The value $p = 2$ corresponds to smooth functions with a continuous gradient at $h = 0$ and $p$ values near 1 or less correspond to less smoothness.

The parameters $\boldsymbol{\omega} = [\omega_1, \omega_2, \ldots, \omega_d]$ ($\omega_k \geq 0$) are called *characteristic length-scales* [45], which are "width parameters that control how far a training point's influence extends" [7]. As shown in Figure 6 (right), given $p = 2$ and at the same distance, larger $\omega$ values result in higher correlation, indicating that output values are close. Smaller $\omega$ values lead to lower correlation, meaning that there will be large difference between outputs even for nearby points.

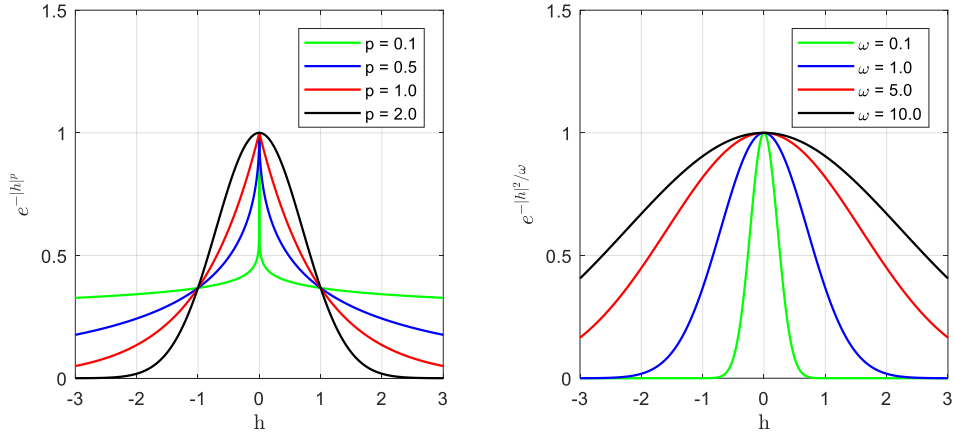

Figure 6: Demonstration of the correlation with different values of roughness parameter $p$ (left) and characteristic length scale $\omega$ (right). The x-axis $h$ means the distance between two points in one-dimension. The left subfigure fixes $\omega = 1$ while the right subfigure fixes $p = 2$.

Table 1. Common spatial correlation kernels.

| Name | Expression ($h = |\mathbf{x}^{(i)} - \mathbf{x}^{(j)}|$) |
| --- | --- |
| Linear | $\mathcal{R}(h) = \max\left(0, 1 - \frac{|h|}{\omega}\right)$ |
| Exponential | $\mathcal{R}(h) = \exp\left(-\frac{|h|}{\omega}\right)$ |
| Power-exponential | $\mathcal{R}(h) = \exp\left(-\left(\frac{|h|}{\omega}\right)^p\right)$ |
| Gaussian | $\mathcal{R}(h) = \exp\left(-\frac{|h|^2}{2\omega^2}\right)$ |
| Matérn $\nu = 3/2$ | $\mathcal{R}(h) = \left(1 + \frac{\sqrt{3}|h|}{\omega}\right)\exp\left(-\frac{\sqrt{3}|h|}{\omega}\right)$ |
| Matérn $\nu = 5/2$ | $\mathcal{R}(h) = \left(1 + \frac{\sqrt{5}|h|}{\omega} + \frac{5h^2}{3\omega^2}\right)\exp\left(-\frac{\sqrt{5}|h|}{\omega}\right)$ |

From the above discussion, apparently the choice of the correlation kernel is crucial to the performance of GP. It is required that the kernel is positive semidefinite. A commonly accepted approach is to select the correlation kernel beforehand from kernels that are known to be positive semidefinite. Table 1 shows some commonly used spatial correlation kernels. Note that the parameter $\omega$ can be different for each of the $d$ dimensions. Among these kernels, Matérn family kernels have been widely used by statisticians, while engineers often favor the Gaussian (or square-exponential) kernel [52]. The latter results in a smooth and infinitely differentiable function which is desirable for many engineering applications. For some other detailed discussions of the correlation kernels, see [45][53][54].

### 3.3. Prediction and Mean Square Error formulas

Given the design matrix $\mathbf{X}$ and the corresponding output values $\mathbf{y}$, to predict the output value at an untried input location $\mathbf{x}^*$ using GP, we first define the following components:

1. The set of regression functions $\mathbf{f} = [f_1, f_2, \ldots, f_n]^T$ evaluated at the untried point $\mathbf{x}^*$:

$$\mathbf{f}(\mathbf{x}^*) = [f_1(\mathbf{x}^*), f_2(\mathbf{x}^*), \ldots, f_n(\mathbf{x}^*)]^T$$

2. The set of regression functions $\mathbf{f} = [f_1, f_2, \ldots, f_n]^T$ evaluated at the m design sites:

$$\mathbf{F} = [\mathbf{f}(\mathbf{x}^{(1)}), \mathbf{f}(\mathbf{x}^{(2)}), \ldots, \mathbf{f}(\mathbf{x}^{(m)})]^T = \begin{bmatrix} f_1(\mathbf{x}^{(1)}) & \cdots & f_n(\mathbf{x}^{(1)}) \\ \vdots & \ddots & \vdots \\ f_1(\mathbf{x}^{(m)}) & \cdots & f_n(\mathbf{x}^{(m)}) \end{bmatrix}$$

3. The correlation of $\mathbf{x}^*$ with the m design sites:

$$\mathbf{r}(\mathbf{x}^*) = \left[\mathcal{R}(\mathbf{x}^*, \mathbf{x}^{(1)}), \mathcal{R}(\mathbf{x}^*, \mathbf{x}^{(2)}), \ldots, \mathcal{R}(\mathbf{x}^*, \mathbf{x}^{(m)})\right]^T$$

4. The correlation between the m design sites:

$$\mathbf{R} = \begin{bmatrix} \mathcal{R}(\mathbf{x}^{(1)}, \mathbf{x}^{(1)}) & \mathcal{R}(\mathbf{x}^{(1)}, \mathbf{x}^{(2)}) & \cdots & \mathcal{R}(\mathbf{x}^{(1)}, \mathbf{x}^{(m)}) \\ \mathcal{R}(\mathbf{x}^{(2)}, \mathbf{x}^{(1)}) & \mathcal{R}(\mathbf{x}^{(2)}, \mathbf{x}^{(2)}) & \cdots & \mathcal{R}(\mathbf{x}^{(2)}, \mathbf{x}^{(m)}) \\ \vdots & \vdots & \ddots & \vdots \\ \mathcal{R}(\mathbf{x}^{(m)}, \mathbf{x}^{(1)}) & \mathcal{R}(\mathbf{x}^{(m)}, \mathbf{x}^{(2)}) & \cdots & \mathcal{R}(\mathbf{x}^{(m)}, \mathbf{x}^{(m)}) \end{bmatrix}$$

Table 2. Symbols used in GP metamodel formulation.

| Symbols | Dimension | Description |
| --- | --- | --- |
| $d$ | $1 \times 1$ | number of input factors |
| $m$ | $1 \times 1$ | number of design points |
| $n$ | $1 \times 1$ | number of basis functions |
| $\boldsymbol{\beta}$ | $n \times 1$ | vector of regression coefficients |
| $\hat{\boldsymbol{\beta}}$ | $n \times 1$ | estimator of $\boldsymbol{\beta}$ |
| $\mathbf{f}$ | $n \times 1$ | vector of regression functions |
| $\mathbf{x}^{(i)}$ | $d \times 1$ | $i$th design point |
| $y_i$ | $1 \times 1$ | output value of $i$th design point |
| $\mathbf{X}$ | $m \times d$ | ensemble of $m$ design points |
| $\mathbf{y}$ | $m \times 1$ | ensemble of output values at $\mathbf{X}$ |
| $\mathbf{x}^*$ | $d \times 1$ | untried input point, to be predicted |
| $\mathbf{f}(\mathbf{x}^*)$ | $n \times 1$ | $\mathbf{f}$ evaluated at untried location $\mathbf{x}^*$ |
| $\mathbf{F}$ | $m \times n$ | $\mathbf{f}$ evaluated at design sites $\mathbf{X}$ |
| $\mathbf{r}(\mathbf{x}^*)$ | $m \times 1$ | correlation between $\mathbf{x}^*$ and design sites $\mathbf{X}$ |
| $\mathbf{R}$ | $m \times m$ | correlation between design sites $\mathbf{X}$ |
| $\hat{y}(\mathbf{x}^*)$ or $\mu_{\hat{y}}(\mathbf{x}^*)$ | $1 \times 1$ | predicted output at $\mathbf{x}^*$ |
| $\mathrm{MSE}[\hat{y}(\mathbf{x}^*)]$ or $\sigma_{\hat{y}}^2(\mathbf{x}^*)$ | $1 \times 1$ | variance of the prediction at $\mathbf{x}^*$ |

The details of all the symbols are summarized in Table 2. The GP prediction of the output at $\mathbf{x}^*$ is given by:

$$\hat{y}(\mathbf{x}^*) = \mu_{\hat{y}}(\mathbf{x}^*) = \mathbf{f}^T(\mathbf{x}^*)\hat{\boldsymbol{\beta}} + \mathbf{r}^T(\mathbf{x}^*)\mathbf{R}^{-1}(\mathbf{y} - \mathbf{F}\hat{\boldsymbol{\beta}}) \quad (14)$$

The estimator for regression coefficients $\hat{\boldsymbol{\beta}}$ is given by the least-squares estimate:

$$\hat{\boldsymbol{\beta}} = (\mathbf{F}^T\mathbf{R}^{-1}\mathbf{F})^{-1}\mathbf{F}^T\mathbf{R}^{-1}\mathbf{y} \quad (15)$$

The predictor in Equation (14) can be proven to be the *Best Linear Unbiased Predictor (BLUP)* of the output [36]. GP also provide the *Mean Square Error* (MSE, or the variance) of the predictor $\hat{y}(\mathbf{x}^*)$:

$$\text{MSE}[\hat{y}(\mathbf{x}^*)] = \sigma_{\hat{y}}^2(\mathbf{x}^*) = \sigma^2 \left[ 1 - [\mathbf{f}^T(\mathbf{x}^*) \quad \mathbf{r}^T(\mathbf{x}^*)] \begin{bmatrix} \mathbf{0} & \mathbf{F}^T \\ \mathbf{F} & \mathbf{R} \end{bmatrix}^{-1} \begin{bmatrix} \mathbf{f}(\mathbf{x}^*) \\ \mathbf{r}(\mathbf{x}^*) \end{bmatrix} \right] \quad (16)$$

Equation (16) can be expanded in another popular form:

$$\begin{aligned} \text{MSE}[\hat{y}(\mathbf{x}^*)] &= \sigma_{\hat{y}}^2(\mathbf{x}^*) \\ &= \sigma^2 \left[ 1 - \mathbf{r}^T(\mathbf{x}^*)\mathbf{R}^{-1}\mathbf{r}(\mathbf{x}^*) + \left(\mathbf{F}^T\mathbf{R}^{-1}\mathbf{r}(\mathbf{x}^*) - \mathbf{f}(\mathbf{x}^*)\right)^T (\mathbf{F}^T\mathbf{R}^{-1}\mathbf{F})^{-1} \left(\mathbf{F}^T\mathbf{R}^{-1}\mathbf{r}(\mathbf{x}^*) - \mathbf{f}(\mathbf{x}^*)\right) \right] \end{aligned} \quad (17)$$

The detailed derivation of Equations (14) - (17) can be found in [55]. Together they formulate the most general case of GP model, named *Universal Kriging (UK)*. *Ordinary Kriging (OK)* is a special case of UK when the basis functions reduce to only $f_1(\mathbf{x}) = 1$ (as the single basis function), which makes the trend to be a constant function whose value is determined by the unknown parameter $\beta_1$. *Simple Kriging (SK)* is the case when the mean function is a known constant value $\mu$:

$$\hat{y}(\mathbf{x}^*) = \mu_{\hat{y},\text{SK}}(\mathbf{x}^*) = \mu + \mathbf{r}^T(\mathbf{x}^*)\mathbf{R}^{-1}(\mathbf{y} - \mu) \quad (18)$$

$$\text{MSE}[\hat{y}(\mathbf{x}^*)] = \sigma_{\hat{y},\text{SK}}^2(\mathbf{x}^*) = \sigma^2[1 - \mathbf{r}^T(\mathbf{x}^*)\mathbf{R}^{-1}\mathbf{r}(\mathbf{x}^*)] \quad (19)$$

The mean functions of UK usually take the form of low-order polynomial regression. If such mean function can correctly capture the trends in the simulation data, then UK tends to give better accuracy than OK and SK. However, in cases when a prior knowledge of the trends of the simulation data is unknown, arbitrary specifications of the mean functions can cause inaccuracies. In fact OK is the most widely used version of GP [52].

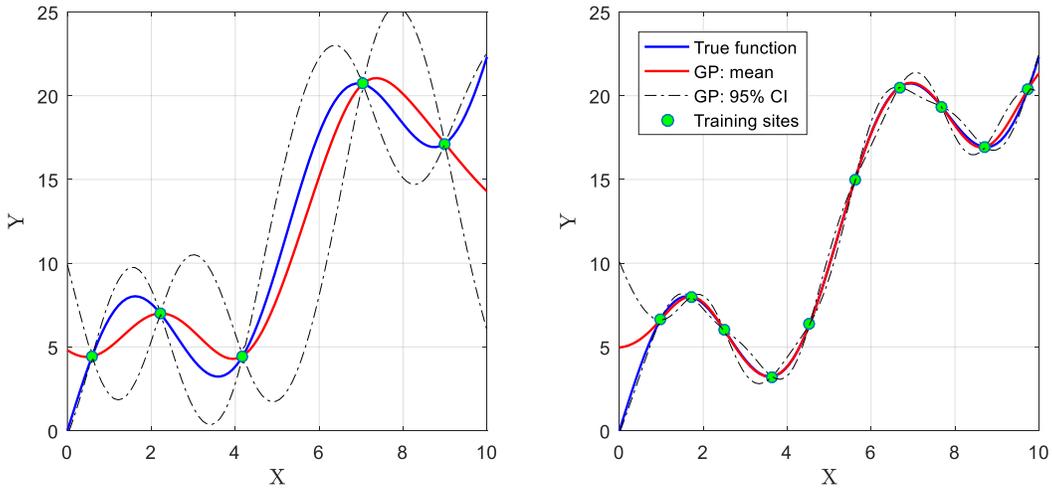

Figure 7: A demonstration of the GP metamodeling with 5 training points (left) and 10 training points (right), based on a simple test function.

Figure 7 demonstrates GP metamodeling results for a simple test function over the range of $[0, 10]$. The true function is only known at 5 and 10 training points respectively. The dash-dot lines correspond to 95% confidence intervals (CIs) of the prediction, which is the interval given by $[\mu_{\hat{y}}(\mathbf{x}^*) - 1.96 * \sigma_{\hat{y}}(\mathbf{x}^*), \mu_{\hat{y}}(\mathbf{x}^*) + 1.96 * \sigma_{\hat{y}}(\mathbf{x}^*)]$. From Figure 7 we can see that:

1) GP metamodel always interpolates the design sites. Detailed derivation of the interpolation property can be found in [55].

2) The MSE of the prediction decreases as the untried point gets closer to training points. At training points the variance of prediction is zero. When the number of training sites increases from 5 to 10, the agreement of the GP mean with true function is greatly improved and the MSEs become much smaller.

Given the mathematical form of GP and the formulas for prediction and MSE, the process of using GP metamodeling involves a few other issues, including "design of computer experiments", "parameter estimation" and "accuracy assessment". There topics are included in the appendices to avoid detour in the flow of the narrative.

"Design of computer experiments" (Appendix A) is the process to select input locations to run the computer code and provide training samples for the GP emulator. Given these training samples, "parameter estimation" (Appendix B) is the process to evaluate all the unknowns in the GP emulator. Parameter estimation aims to determine the GP parameter values that lead to the best predictive performance. There are $(n + 2d + 1)$ unknowns in the GP emulator:

1) $n$ regression coefficients $\boldsymbol{\beta}$;

2) $d$ roughness parameters $\mathbf{p}$; Note that these parameters are usually known once we have chosen a certain correlation kernel. For example, for Gaussian kernels these parameters will all be 2;

3) $d$ characteristic length scale parameters $\boldsymbol{\omega}$;

4) One process variance parameter $\sigma^2$. Note that many researchers, instead of the variance parameter, use the term "precision" defined as $\lambda = 1/\sigma^2$ [24][25][31][33].

Together, these $(n + 2d + 1)$ unknown parameters $\boldsymbol{\Psi} = \{\boldsymbol{\beta}, \sigma^2, \boldsymbol{\omega}, \mathbf{p}\}$ are called *hyperparameters*, whose values can be evaluated either by Maximum Likelihood Estimation (MLE) or Cross Validation (CV). Detailed derivations and comparison of both approaches can be found in Appendix B. Finally, "accuracy assessment" (Appendix C) is the process to evaluate the capability of the GP emulator to reproduce the computer code before it can serve as a "surrogate" of the computer code. In Appendix D, we provide some discussions on the implementation issues of GP.

## 4. Full and Modular Bayesian Approaches

In this section, we introduce the full and modular Bayesian approaches that have been used in the literature for Bayesian calibration. The majority of their implementations follow the seminal work of Kennedy and O'Hagan [14], hereafter referred to as the "KOH" method. The KOH method became part of the later developed framework called Bayesian Analysis of Computer Code Outputs (BACCO) [44]. Research in DACE (Design and Analysis of Computer Experiments) [35][36] was the forerunner of BACCO. The BACCO approach uses GP to build emulators for complex computer models, and then use these emulators to perform calibration, uncertainty and sensitivity analysis.

Bayesian analysis simultaneously incorporates all relevant information and deals with all uncertainties in the model. In full Bayesian approach, all the unknown hyperparameters in the GP emulators for both the computer model and its discrepancy are treated in a similar way as the calibration parameters. They are assigned priors which also enter the likelihood function. Eventually posteriors of GP hyperparameters and calibration parameters are solved for all together. Full Bayesian results in an extremely complicated function for the posteriors of calibration parameters and GP hyperparameters. Then the GP hyperparameters need to be integrated out from the joint posterior to get marginal distributions of the calibration parameters [56]. However, in modular Bayesian approach, the estimation of calibration parameters, GP hyperparameters for computer model and model discrepancy are all separated. The details of both approaches will be described in the following sections. We will use $\boldsymbol{\theta}$ instead of $\boldsymbol{\theta}^*$ to represent the target of inverse UQ for notational simplicity.

### 4.1. Full Bayesian approach

The key components of full Bayesian approach are illustrated in Figure 8. Given experimental observations, a test source allocation (TSA) algorithm is first implemented to separate them for inverse UQ $y^E(x^{IUQ})$ and validation $y^E(x^{VAL})$ because the same data should not be used for both purposes. Next two GP emulators are built for the computer model and its discrepancy:

$$y^M(x, \theta) \sim \mathcal{GP}\{(f^M(x, \theta))^T \beta^M, \sigma_M^2 \mathcal{R}^M \langle (x^{(i)}, \theta^{(i)}), (x^{(j)}, \theta^{(j)}) \rangle\} \quad (20)$$

$$\delta(x) \sim \mathcal{GP}\{(f^\delta(x))^T \beta^\delta, \sigma_\delta^2 \mathcal{R}^\delta \langle x^{(i)}, x^{(j)} \rangle\} \quad (21)$$

Note that the first GP emulator takes both $x$ and $\theta$ as inputs, while the second GP emulator only takes $x$. The basis functions $f^M$ and $f^\delta$ are chosen by the user. The hyperparameters for the computer model GP emulator $\Psi^M = \{\beta^M, \sigma_M^2, \omega^M, p^M\}$ and the model discrepancy GP emulator $\Psi^\delta = \{\beta^\delta, \sigma_\delta^2, \omega^\delta, p^\delta\}$ are unknown. Replacing $y^M(x, \theta)$ and $\delta(x)$ in the "model updating equation" we get:

$$y^E = \mathcal{GP}\{(f^M)^T \beta^M, \sigma_M^2 \mathcal{R}^M\} + \mathcal{GP}\{(f^\delta)^T \beta^\delta, \sigma_\delta^2 \mathcal{R}^\delta\} + \varepsilon \quad (22)$$

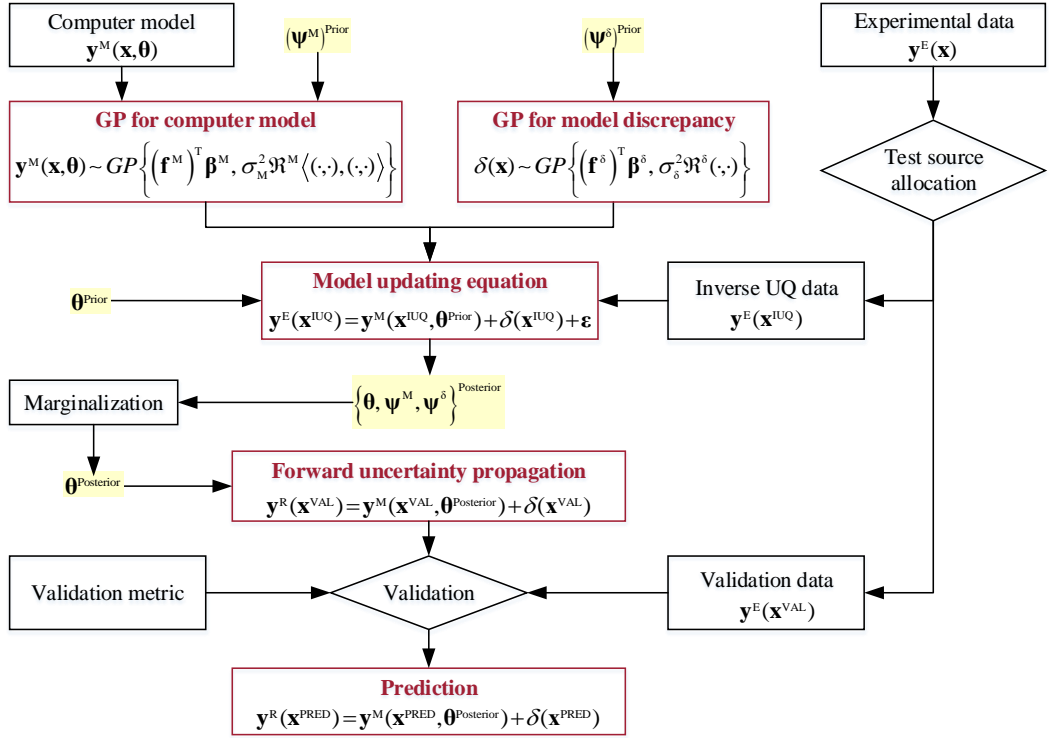

Figure 8: Workflow of the full Bayesian approach.

Following the KOH approach, $y^M(x, \theta)$, $\delta(x)$ and $\varepsilon$ are assumed independent. Such assumption has been adopted in all the previous work on Bayesian calibration and significantly simplifies the calculations. Full Bayesian approach then proceeds with assigning prior distributions for all the unknowns shown below:

$$\{\theta, \Psi^M, \Psi^\delta, \sigma_\varepsilon^2\} = \{\theta, \beta^M, \sigma_M^2, \omega^M, p^M, \beta^\delta, \sigma_\delta^2, \omega^\delta, p^\delta, \sigma_{exp}^2\} \quad (23)$$

Note that the measurement error $\sigma_{exp}^2$ is sometimes treated as known [27] because the error rates for most instrumentation tools are known. In this work, we assume that this parameter is in fact reported with the benchmark data. If the $\sigma_{\text{exp}}^2$ is unknown, it can be learnt simultaneously together with the calibration parameters and GP hyperparameters [24][25][31]. Furthermore, upon selection of a correlation kernel, the roughness parameters $\mathbf{p}^M$ and $\mathbf{p}^\delta$ are usually known. However, even after assuming they are known, we are still left with many parameters to evaluate $\{\boldsymbol{\theta}, \boldsymbol{\Psi}^M, \boldsymbol{\Psi}^\delta\} = \{\boldsymbol{\theta}, \boldsymbol{\beta}^M, \sigma_M^2, \boldsymbol{\omega}^M, \boldsymbol{\beta}^\delta, \sigma_\delta^2, \boldsymbol{\omega}^\delta\}$. There is no guidance for the selection of proper prior distributions for them, especially the hyperparameters. Previously, researchers have used distributions like normal, gamma, beta and inverse gamma for these hyperparameters [24][25][31], however with no justification about why they chose those specific distributions.

After the selection of priors, next a joint likelihood function is formulated for all the unknowns $\{\boldsymbol{\theta}, \boldsymbol{\Psi}^M, \boldsymbol{\Psi}^\delta\}$. Given the inverse UQ measurement data $\boldsymbol{y}^E(\mathbf{x}^{\text{IUQ}})$ (i.e., the measurement data obtained at design variables $\mathbf{x}^{\text{IUQ}}$) and computer model training samples $\boldsymbol{y}^M(\mathbf{x}^{\text{IUQ}}, \boldsymbol{\theta}^{\text{train}})$ where $\boldsymbol{\theta}^{\text{train}}$ represents the training parameter values, MCMC sampling can be used to generate samples for the unknowns. In this way, full Bayesian analysis achieves the uncertainties in calibration parameters and model discrepancy term simultaneously. Marginalization is required to integrate out $\{\boldsymbol{\Psi}^M, \boldsymbol{\Psi}^\delta\}$ to get $\boldsymbol{\theta}^{\text{Posterior}}$. After inverse UQ, $\boldsymbol{\theta}^{\text{Posterior}}$ can be directly used for validation and prediction, two tasks that are not the focus of this paper.

### 4.2. Modular Bayesian approach

Modularization is a technique to separate various modules in Bayesian analysis to prevent suspect information belonging to one part from overly influencing another part [33]. The most important difference of modular Bayesian with full Bayesian is that the former separates the estimation of $\{\boldsymbol{\theta}, \boldsymbol{\Psi}^M, \boldsymbol{\Psi}^\delta\}$. Modular Bayesian uses plausible estimates (e.g. MLEs) of $\{\boldsymbol{\Psi}^M, \boldsymbol{\Psi}^\delta\}$ and treat them as if they were the true values of $\{\boldsymbol{\Psi}^M, \boldsymbol{\Psi}^\delta\}$ [14]. Figure 9 shows the detailed flowchart of the modular Bayesian approach, which includes the following major distinctions with full Bayesian approach:

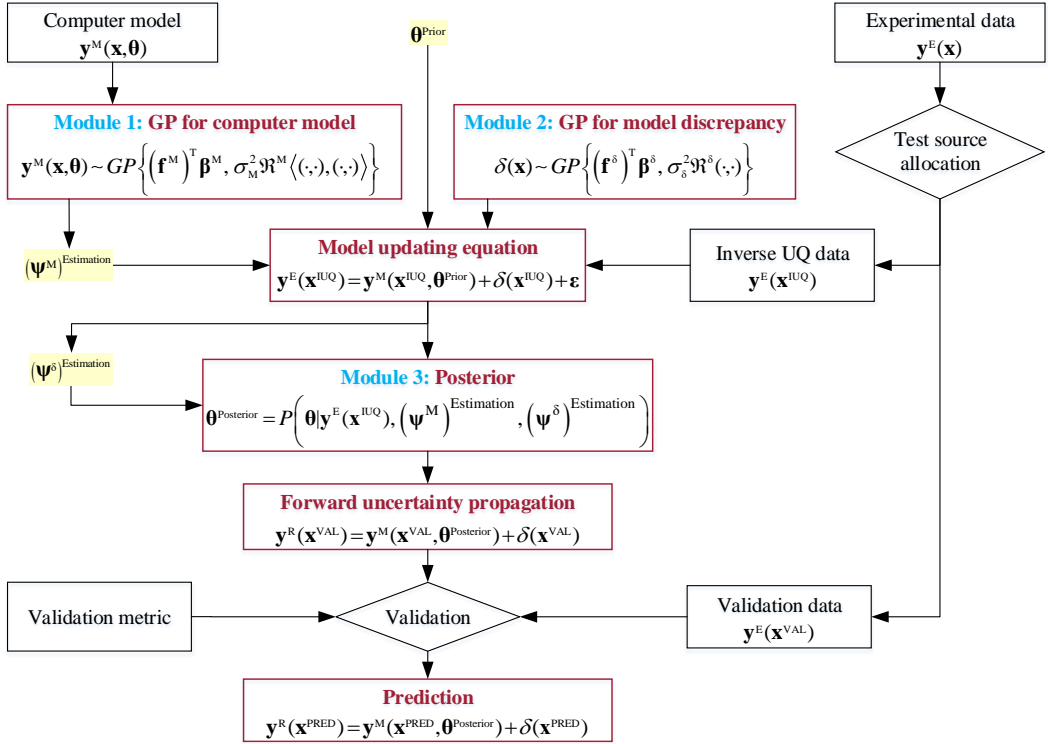

Figure 9: Workflow of the modular Bayesian approach.

1) Module 1 replaces the computer code $\boldsymbol{y}^M$ with a GP emulator, whose hyperparameters $\boldsymbol{\Psi}^M$ is deterministically estimated based on the computational results at the training sites $\boldsymbol{y}^M(\mathbf{x}^{\text{IUQ}}, \boldsymbol{\theta}^{\text{train}})$. The estimation process of $\mathbf{\Psi}^M$ is straightforward using the parameter estimation methods mentioned in Appendix B, such as MLE.

2) Module 2 fits a second GP emulator to the model discrepancy term $\delta(\mathbf{x})$. The hyperparameters $\mathbf{\Psi}^\delta$ are deterministically estimated given the GP emulator from module 1, the measurement data $\mathbf{y}^E(\mathbf{x}^{IUQ})$ and the prior distributions of calibration parameters $\mathbf{\theta}^{Prior}$. Starting from the "model updating equation", one forms a likelihood function for $\mathbf{y}^E(\mathbf{x}^{IUQ})$ based on module 1 and $\mathbf{\theta}^{Prior}$ and then evaluates the MLEs of $\mathbf{\Psi}^\delta$. Closed forms of the likelihood function is possible using Gaussian correlation kernel, constant basis functions and normal [14] or uniform [23] priors for $\mathbf{\theta}^{Prior}$.

3) Once module 1 and module 2 become known (GP hyperparameters have been solved), module 3 achieves the posterior $\mathbf{\theta}^{Posterior}$ from the likelihood function through MCMC sampling. Note that unlike the full Bayesian approach in which marginalization is required, here $\mathbf{\theta}^{Posterior}$ is conditioned on the (deterministic) estimations of the GP hyperparameters $\{\mathbf{\Psi}^M, \mathbf{\Psi}^\delta\}$. However, this is also a limitation of the modular Bayesian approach, because the uncertainties in $\{\mathbf{\Psi}^M, \mathbf{\Psi}^\delta\}$ are ignored. This is why such a method is only "empirical or partial" Bayesian [31], as opposed to the "full" Bayesian approach of the previous section.

Note that a single definition of modules, unanimously agreed upon in literature, does not exist. For example, [33] chose the (1) computer model output $\mathbf{y}^M$, (2) the model discrepancy $\delta$ and (3) the field data $\mathbf{y}^E$ as three modules, while in [23], the (1) computer model output $\mathbf{y}^M$, (2) model discrepancy $\delta$, (3) posterior $\mathbf{\theta}^{Posterior}$ and (4) prediction of $\mathbf{y}^E$ and $\delta$ were treated as four modules. In the current work we use a definition similar to [23] as shown in Figure 9.

### 4.3. Comparison of full vs modular Bayesian approaches

Full Bayesian analysis is theoretically superior, but computationally intractable. This has motivated a larger body of research invested on the modular Bayesian approach [23][27][30][31][32][33], compared to the limited works focusing on full Bayesian [24][25]. Full Bayesian requires that the user have reasonably good priors for the GP hyperparameters $\{\mathbf{\Psi}^M, \mathbf{\Psi}^\delta\}$. This is very difficult in practice especially for the model discrepancy term $\{\mathbf{\Psi}^\delta\}$. The joint posterior for all the unknowns shown in Equation (23) can be high dimensional, posing challenges for MCMC sampling and subsequent marginalization.

Also, even though the full Bayesian approach does take into account the uncertainties in the hyperparameters, for the calibration-validation-prediction process, the uncertainties in $\mathbf{\theta}$ and $\mathbf{\Psi}^\delta$ tend to overwhelm the uncertainties in computer model approximations (i.e. $\mathbf{\Psi}^M$). It was found that using MLE plug-in (i.e. modular Bayesian) typically leads to similar results with those from using full Bayesian. Therefore, researchers recommended using the partial/empirical/modular Bayesian rather than full Bayesian [31].

Liu et al. [33] also discussed other advantages of modularization. For example, separating good modules from suspect modules can avoid information "contamination". Suspect modules are modules that are unknown or improperly specified (especially when there is no better choice). Moreover, modularization can also (1) reduce the computational complexity by reasonable simplification, (2) make scientific understanding and development more convenient, (3) improve mixing and convergence of MCMC sampling.

### 4.4. Discussions on the modeling of model discrepancy term

In Section 2.2, we have briefly talked about the model discrepancy term $\delta(\mathbf{x})$. Model discrepancy accounts for inadequacies that are caused by missing or insufficient physics and numerical approximations built into the simulation model, which leads to systematic differences between the simulation model and the reality [25]. Ignoring model discrepancy during inverse UQ may cause "over-fitting" and subsequent prediction errors. However, due to the inherent difficulty in the mathematical description of the model discrepancy or because of limited amount of data, the model discrepancy term was simply ignored in older applications of Bayesian calibration [10] [11] [15] [19] [20] [21] [22] [26][43]. However, it has been considered in recent studies, such as [14][23][24][25][27].

The primary challenge in dealing with model discrepancy is that this discrepancy cannot be directly observed or measured. Another difficulty is that model discrepancy is almost always seriously "confounded" with other model uncertainties, such as the parameter uncertainties [33]. All the past studies on the full or modular Bayesian approaches provide mathematical descriptions of the model discrepancy term. Such a choice has been proven successful in many applications as shown in the aforementioned publications. However, an important issue associated with these approaches is whether this model discrepancy representation can be extrapolated to various validation and prediction conditions.

As shown in Figure 8 and 9, after learning about $\boldsymbol{\theta}^{\text{Posterior}}$, the realities in the validation and prediction domain are described as:

$$y^{\text{R}}(\mathbf{x}^{\text{VAL}}) = y^{\text{M}}(\mathbf{x}^{\text{VAL}}, \boldsymbol{\theta}^{\text{Posterior}}) + \delta(\mathbf{x}^{\text{VAL}}) \tag{24}$$

$$y^{\text{R}}(\mathbf{x}^{\text{PRED}}) = y^{\text{M}}(\mathbf{x}^{\text{PRED}}, \boldsymbol{\theta}^{\text{Posterior}}) + \delta(\mathbf{x}^{\text{PRED}}) \tag{25}$$

Where $\mathbf{x}^{\text{VAL}}$ and $\mathbf{x}^{\text{PRED}}$ stand for the designs variables in the validation and prediction domain respectively. The first term on the right hand sides, i.e. computer model output $y^{\text{M}}$, can be either the computer model output itself or its GP emulator, while the second term $\delta$ is always the GP emulator. Because the information acquired about the model discrepancy hyperparameters $\left(\boldsymbol{\Psi}^{\delta}\right)^{\text{Posterior}}$ (Figure 8) or $\left(\boldsymbol{\Psi}^{\delta}\right)^{\text{Estimation}}$ (Figure 9) is based on model simulation $y^{\text{M}}(\mathbf{x}^{\text{IUQ}}, \boldsymbol{\theta}^{\text{train}})$ and measurement data $y^{\text{E}}(\mathbf{x}^{\text{IUQ}})$ in the inverse UQ domain, Equation (24) and (25) essentially involve extrapolation of such information to the validation and prediction conditions.

Inverse UQ with both full and modular Bayesian are fully data-driven. Therefore, these methods should be used with great caution. Extrapolation outside the range of the inverse UQ domain is questionable. As discussed in [25], the quality of such extrapolation largely depends on the reliability of the model discrepancy term. What we have learnt about the model discrepancy in the inverse UQ domain may not be applicable to the validation and prediction domain. See [24] for an example in which the model discrepancy is large in magnitude, but the posterior $\boldsymbol{\theta}^{\text{Posterior}}$ is similar with that obtained when the model discrepancy is zero. Apparently adding a large discrepancy value when the "true" discrepancy is zero causes large error in the validation/prediction domains.

Besides the reliability of the model discrepancy term, extrapolation using GP emulator is inherently dangerous. From Figure 7 we can see that GP emulator can have large mean prediction errors and significant MSE outside of the training domain. Even though the full and modular Bayesian methodologies outlined in Section 4.1 and Section 4.2 have been accepted for a long time, we recommend making some improvement in describing the model discrepancy term to avoid extrapolation as discussed in Section 5.

## 5. An Improved Modular Bayesian Approach

An improved modular Bayesian approach is developed and described in this section. It is much more straightforward to understand and apply, and it is capable of avoiding the issue of extrapolation of the GP emulators. We also use a GP emulator to represent the model discrepancy term $\delta(\mathbf{x}) \sim \mathcal{GP}\left\{\left(\mathbf{f}^{\delta}\right)^{\text{T}} \boldsymbol{\beta}^{\delta}, \sigma_{\delta}^{2} \mathcal{R}^{\delta}\right\}$. To solve for the GP hyperparameters $\boldsymbol{\Psi}^{\delta} = \{\boldsymbol{\beta}^{\delta}, \sigma_{\delta}^{2}, \boldsymbol{\omega}^{\delta}, \mathbf{p}^{\delta}\}$ with MLE, we need training data whose input is $\mathbf{x}$ and output is the computer code simulation error (recall that model discrepancy is also called model error). Because there are no direct observations of $\delta(\mathbf{x})$, we need substitutes of such "observation data" for simulation error. Three intuitively natural modularization schemes were compared in [33] to fix $\boldsymbol{\Psi}^{\delta}$.

- Modular approach 1: Treat the differences between measurement data and model simulations (run at prior means or nominal values of $\boldsymbol{\theta}$, but use the same design variables with measurement data) as the realizations of model discrepancy. Then $\boldsymbol{\Psi}^{\delta}$ can be estimated with MLE.

- Modular approach 2: Sample the calibration parameters from $\boldsymbol{\theta}^{\text{Prior}}$. Then sample $\boldsymbol{\Psi}^{\delta}$ from their posteriors, which is generated conditioning on the samples of $\boldsymbol{\theta}^{\text{Prior}}$. The posterior sample mean will be used as the fixed values of $\boldsymbol{\Psi}^{\delta}$.

- Modular approach 3: Initially assume the model is perfect (model discrepancy is zero). Then solve for $\boldsymbol{\theta}^{\text{Posterior}}$ based on this assumption. The resulting posterior will be used as a "new" prior and proceed with modular approach 1.

Modular approach 1 was demonstrated to have the best performance for the simple test problem in [33]. Actually such treatment has been used in other related research, for example design-driven validation [57] and Bayesian validation [58]. In the examples of [58], it was shown that the model discrepancy posterior using the modular Bayesian approach described in Section 4.2 is the same as that obtained by fitting a single GP to $\{y^E - y^M\}$. Whether this conclusion applies to a more complicated real problem is unknown and has not been investigated.

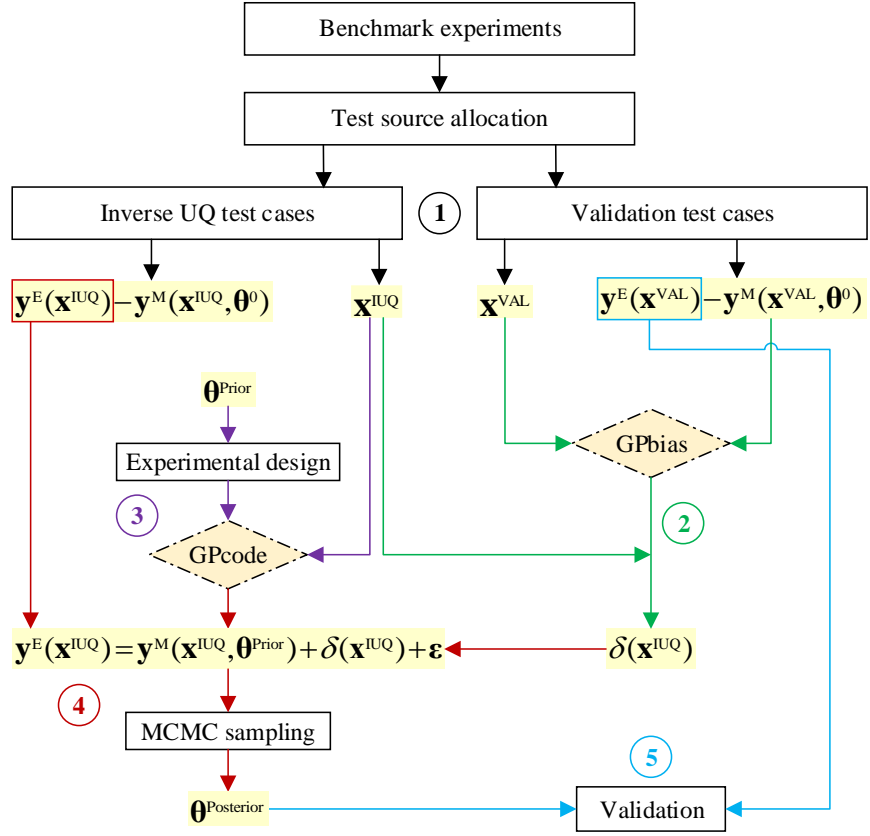

Figure 10: Flowchart of the improved modular Bayesian approach

The revised modular Bayesian approach takes the modular approach 1 and its main components are shown in Figure 10, which consists of the following steps:

- Step 1: separate the measurement data for inverse UQ and validation:

  This step is shown as black arrows in Figure 10. Instead of random selection, we use a carefully designed sequential algorithm for TSA, which will be discussed in our companion paper [29]. For now, we assume that the given tests have already been properly separated.

- Step 2: mathematical description of the model discrepancy term $\delta(\mathbf{x})$:

  This step is shown as green arrows in Figure 10. The computer code $y^M(\mathbf{x}, \boldsymbol{\theta})$ is first executed at the input settings of validation domain $\mathbf{x}^{VAL}$, with the $\boldsymbol{\theta}$ fixed at nominal values or prior mean values $\boldsymbol{\theta}^0$ which can be viewed as our current best knowledge of $\boldsymbol{\theta}$. The resulting simulations are denoted as $y^M(\mathbf{x}^{VAL}, \boldsymbol{\theta}^0)$. Then $\mathbf{x}^{VAL}$ and $\{y^E(\mathbf{x}^{VAL}) - y^M(\mathbf{x}^{VAL}, \boldsymbol{\theta}^0)\}$ are used as training inputs and output respectively to fit a GP emulator called "GPbias". Note that the fitting of "GPbias" uses the observations noises of $y^E(\mathbf{x}^{VAL})$ as a "nugget" term to reflect the uncertainty in the data. Evaluating "GPbias" at $\mathbf{x}^{IUQ}$ results in an estimation of the model discrepancy term $\delta(\mathbf{x}^{IUQ})$, which will enter the likelihood function during MCMC sampling.

- Step 3: fit a GP emulator for the computer code:

  This step is shown as purple arrows in Figure 10. Another GP emulator called "GPcode" is fitted to replace the computer code during MCMC sampling. "GPcode" uses $(\mathbf{x}^{IUQ}, \boldsymbol{\theta}^{train})$ as training inputs and $y^M(\mathbf{x}^{IUQ}, \boldsymbol{\theta}^{train})$ as training outputs, where $\boldsymbol{\theta}^{train}$ is an experimental design of $\boldsymbol{\theta}$ following $\boldsymbol{\theta}^{Prior}$.

- Step 4: MCMC sampling:

    This step is shown as red arrows in Figure 10. "GPcode", $\delta(\mathbf{x}^{\text{IUQ}})$ and $\mathbf{y}^{\text{E}}(\mathbf{x}^{\text{IUQ}})$ enter the posterior PDF which is then explored with MCMC sampling to obtain $\mathbf{\theta}^{\text{Posterior}}$.

- Step 5: validation of $\mathbf{\theta}^{\text{Posterior}}$:

    This step is shown as light blue arrows in Figure 10. Computer code simulations $\mathbf{y}^{\text{M}}(\mathbf{x}^{\text{VAL}}, \mathbf{\theta}^{\text{Posterior}})$ and $\mathbf{y}^{\text{E}}(\mathbf{x}^{\text{VAL}})$ are compared for validation of the achieved posterior with certain validation metrics.

In our companion paper [29], we will provide a step-by-step demonstration of the application of this improved modular Bayesian approach to a practical problem in nuclear engineering. Note that in the validation domain (and future prediction domain as well), we use $\mathbf{y}^{\text{R}}(\mathbf{x}^{\text{VAL}}) = \mathbf{y}^{\text{M}}(\mathbf{x}^{\text{VAL}}, \mathbf{\theta}^{\text{Posterior}})$ instead of $\mathbf{y}^{\text{R}}(\mathbf{x}^{\text{VAL}}) = \mathbf{y}^{\text{M}}(\mathbf{x}^{\text{VAL}}, \mathbf{\theta}^{\text{Posterior}}) + \delta(\mathbf{x}^{\text{VAL}})$ to avoid extrapolation. Because of the specially designed TSA process in step 1 (see details in our companion paper [29]) and treatment of model discrepancy in step 2, $\mathbf{\theta}^{\text{Posterior}}$ is already informed by the experimental data in the validation domain so that the model discrepancy term $\delta(\mathbf{x}^{\text{VAL}})$ is no longer needed to correct the model simulation $\mathbf{y}^{\text{M}}(\mathbf{x}^{\text{VAL}}, \mathbf{\theta}^{\text{Posterior}})$.

## 6. Conclusions

In nuclear reactor system design and safety analysis, the BEPU methodology requires that output uncertainties to be quantified in order to ensure that the investigated design remains within acceptance criteria. "Expert opinion" and "user self-evaluation" have been widely used to provide input uncertainties in previous uncertainty, sensitivity and validation studies. Inverse UQ, which is a process to more precisely quantify input uncertainties using available experimental data, can replace such ad-hoc expert judgment in the uncertainty characterization. Specifically, inverse UQ identifies the statistical characterization of uncertain input parameters that is most consistent with the available measurement data, and effectively offers information beyond the best-fit values.

In this paper, we used Bayesian analysis to establish the inverse UQ formulation, with systematic and rigorously derived surrogate models constructed by GP. We provided a detailed introduction and comparison of the full and modular Bayesian approaches for inverse UQ. Inverse UQ with both approaches are fully data-driven. Therefore, these methods should be used with great caution. The extrapolation of the model discrepancy term depends on its error structure in the validation/prediction domain. Since the model discrepancy term is trained based on experimental data in the inverse UQ domain, its extrapolation to the validation/prediction domain is a sophisticated and unresolved issue. We proposed an improved modular Bayesian approach that can avoid extrapolating the model discrepancy term. The improved approach is organized in a structure such that the posteriors achieved with data in inverse UQ domain is informed by data in the validation domain. Therefore, over-fitting can be avoided while extrapolation is not required.

In a companion paper [29], the improved modular Bayesian approach will be applied to the nuclear reactor system thermal-hydraulics code TRACE based on void fraction measurement data. The application will demonstrate the ability of our treatment of the model discrepancy to avoid over-fitting. Note that, in this work, we only consider problems with low-to-moderate-dimensional QoIs. When the model output is high-dimensional (e.g. time-dependent), dimensionality reduction is required to speed up the computations required to explore the posterior. See [25][43] for examples on using basis representations (e.g. principal components) for time series outputs.

**Appendix A. Design of Computer Experiments**

As illustrated in Figure 7, a dense design will produce a GP metamodel that has better predictive capability. However, this is in contradiction with the purpose of using GP as metamodel, which is to greatly reduce the computation cost. A good balance can be achieved by improving the selection of the design sites in a process called "*experimental design*" or "*design of computer experiments*" [35][36]. Simple or crude Monte Carlo (MC) sampling [59] selects samples randomly according to the pre-specified probability distributions. Simple MC sampling generally do not have a good coverage of the sampling space [59]. Consequently, a number of sampling methods were developed to promote space-filling sample designs. Latin Hypercube Sampling (LHS) and low discrepancy sequences are two of these sampling approaches.

LHS [60][61] is one of the stratified sampling techniques, which divide range each input range into $N_{LHS}$ segments of equal probability, where $N_{LHS}$ is the number of desired samples. The relative lengths of the segments are determined by the nature of the prescribed PDF. For instance, the segment lengths will be equal for a uniform distribution. LHS selects one sample randomly from each of the segments, resulting in $N_{LHS}$ samples for each input. Based on a pre-specified correlation structure, these samples are then combined in a shuffling operation to create a set of $N_{LHS}$ multi-dimensional samples. Consequently, an obvious feature of resulting sample set is that every row or column has exactly one sample in the hypercube of partitions. In spite of the advantage of LHS in requiring fewer samples, compared to simple MC sampling, to achieve the same statistical accuracy, it can still be prohibitively expensive. To further improve it, a popular variation of LHS, namely the "maximin LHS design" [62][63], selects the samples with the largest minimum distance among multiple designs to achieve a good coverage of the sampling space.

Low-discrepancy sequences, also called quasi-random or sub-random sequences [64][65], are "less random" than a crude MC sequence. It contains samples that tend to be located "more uniformly" than crude MC. Here discrepancy refers to the non-uniformity of the samples within the unit hypercube. Low discrepancy sequences have an advantage over LHS design of being generated sequentially, which is useful when additional samples are sequentially added. They are widely used for approximation of integrals in higher dimensions or for global optimization because superior convergence can be achieved. Examples of low-discrepancy sequences are Sobol sequence [66][67] and Halton sequence [64][65].

By looking at the MSE we directly obtain a measure of the local error for any untried input location. This measure can be used to detect regions of low prediction accuracy, which can provide guidance to enrich the current experimental design. An overall increase in the GP metamodel prediction accuracy can be achieved by placing more samples to the region with large prediction variance. This procedure is referred to as "*adaptive experimental design*" [71].

**Appendix B. Parameter Estimation**

**Maximum Likelihood Estimation (MLE)**

Based on the fundamental assumption of GP, we know that the outputs values **y** of the $m$ design sites follow a joint Gaussian distribution:

$$\mathbf{y}|\mathbf{\Psi} = \mathbf{y}|\{\boldsymbol{\beta}, \sigma^2, \boldsymbol{\omega}, \mathbf{p}\} \sim \mathcal{N}(\mathbf{F}\boldsymbol{\beta}, \sigma^2 \mathbf{R}) \tag{26}$$

where the correlation matrix **R** depends on $\{\boldsymbol{\omega}, \mathbf{p}\}$. The likelihood function can be written as:

$$\mathcal{L}(\mathbf{y}|\{\boldsymbol{\beta}, \sigma^2, \boldsymbol{\omega}, \mathbf{p}\}) = \frac{1}{(\sqrt{2\pi}\sigma)^m \sqrt{|\mathbf{R}|}} \exp\left[-\frac{(\mathbf{y} - \mathbf{F}\boldsymbol{\beta})^T \mathbf{R}^{-1}(\mathbf{y} - \mathbf{F}\boldsymbol{\beta})}{2\sigma^2}\right] \tag{27}$$

The idea behind MLE is to find the set of hyperparameters **Ψ** that maximize the likelihood of the observations **y**. We first assume that the correlation kernel parameters $\{\boldsymbol{\omega}, \mathbf{p}\}$ are known, based on which we solve for $\{\boldsymbol{\beta}, \sigma^2\}$ in closed forms. The regression coefficients are estimated by the generalized least-squares methods and the Best Linear Unbiased Estimator (BLUE) for **β**, as shown in Equation (15), is used.

$$\widehat{\boldsymbol{\beta}}|\{\boldsymbol{\omega}, \mathbf{p}\} = \widehat{\boldsymbol{\beta}}(\boldsymbol{\omega}, \mathbf{p}) = (\mathbf{F}^T \mathbf{R}^{-1} \mathbf{F})^{-1} \mathbf{F}^T \mathbf{R}^{-1} \mathbf{y} \tag{28}$$

Maximizing the likelihood is equivalent to minimizing the negative log-likelihood, which is:

$$-\log\{\mathcal{L}(\mathbf{y}|\mathbf{\Psi})\} = \frac{m}{2}\log(2\pi) + \frac{1}{2}\log(|\mathbf{R}|) + m\log(\sigma) + \frac{(\mathbf{y}-\mathbf{F}\widehat{\boldsymbol{\beta}})^{\mathrm{T}}\mathbf{R}^{-1}(\mathbf{y}-\mathbf{F}\widehat{\boldsymbol{\beta}})}{2\sigma^2} \tag{29}$$

The first two terms are constants as we have assumed that $\{\boldsymbol{\omega}, \mathbf{p}\}$ are known. Taking derivative of the last two terms and we can find the estimator for $\sigma^2$:

$$\widehat{\sigma^2}|\{\boldsymbol{\omega},\mathbf{p}\} = \frac{1}{m}(\mathbf{y}-\mathbf{F}\widehat{\boldsymbol{\beta}})^{\mathrm{T}}\mathbf{R}^{-1}(\mathbf{y}-\mathbf{F}\widehat{\boldsymbol{\beta}}) \tag{30}$$

If we substitute the estimators $\widehat{\sigma^2}$ and $\widehat{\boldsymbol{\beta}}$ into the likelihood function, we get the so called "concentrated likelihood function" [51]. Here we choose the negative log-likelihood function for convenience:

$$-\log\{\mathcal{L}(\mathbf{y}|\mathbf{\Psi})\} = \frac{m}{2}\log(2\pi) + \frac{1}{2}\log(|\mathbf{R}|) + \frac{m}{2}\log(\widehat{\sigma^2}) + \frac{m}{2} = \frac{m}{2}\log(2\pi\widehat{\sigma^2}) + \frac{1}{2}\log(|\mathbf{R}|) + \frac{m}{2} \tag{31}$$

The MLE estimator of $\{\boldsymbol{\omega}, \mathbf{p}\}$ can be achieved solving the following optimization problem:

$$\{\widehat{\boldsymbol{\omega}},\widehat{\mathbf{p}}\} = \arg\min_{\mathcal{D}\{\boldsymbol{\omega},\mathbf{p}\}}\left(\frac{m}{2}\log(2\pi\widehat{\sigma^2}) + \frac{1}{2}\log(|\mathbf{R}|) + \frac{m}{2}\right) \tag{32}$$

Where $\mathcal{D}\{\boldsymbol{\omega}, \mathbf{p}\}$ represents the domain for possible values of $\{\boldsymbol{\omega}, \mathbf{p}\}$. Note that this process can be made easier if we fix the values of $\mathbf{p}$ by choosing correlation kernels beforehand.

**Cross Validation (CV)**

The basic idea of CV is to leave out one or few design sites and their corresponding output(s), train the GP model using the remaining points, and then compare the model predictions and observed outputs at those left out sites. An optimization problem is designed to minimize such prediction error which typically requires looping over all the design sites. Consider a general case of "*K-fold CV*" which divides the domain $\mathcal{D}(\mathbf{X})$ of the design sites $\mathbf{X}$ into K mutually exclusive and collectively exhaustive subsets $\{\mathcal{D}_k, k = 1,2,\ldots, K\}$ (usually with equal size) such that:

$$\mathcal{D}_i \cap \mathcal{D}_j = \emptyset, \quad \forall (i,j) \in \{1,2,\ldots,K\}^2$$

$$\cup_{k=1}^{K} \mathcal{D}_k = \mathcal{D}(\mathbf{X})$$

Suppose that the $k^{\text{th}}$ set of design sites are left out and a GP emulator is trained using the remaining $(K - 1)$ sets. We then make predictions using this emulator on the (left out) $k^{\text{th}}$ set. Such predictions are called "cross-validated predictions" and denoted by $\mu_{\widehat{y},(-k)}(\mathbf{x}^{(k)})$. We repeat the process with $k = \{1,2,\ldots, K\}$ and then minimizing certain objective function to get estimation of the hyperparameters. A common choice of the objective function is given by:

$$\sum_{k=1}^{K}\left[y^{\mathrm{M}}(\mathbf{x}^{(k)}) - \mu_{\widehat{y},(-k)}(\mathbf{x}^{(k)})\right]^2$$

The CV estimate of $\{\boldsymbol{\omega}, \mathbf{p}\}$ can be achieved by minimizing the above objective function:

$$\{\widehat{\boldsymbol{\omega}},\widehat{\mathbf{p}}\} = \arg\min_{\mathcal{D}\{\boldsymbol{\omega},\mathbf{p}\}}\left(\sum_{k=1}^{K}\left[y^{\mathrm{M}}(\mathbf{x}^{(k)}) - \mu_{\widehat{y},(-k)}(\mathbf{x}^{(k)})\right]^2\right) \tag{33}$$

The CV estimate of $\sigma^2$ is calculated using:

$$\widehat{\sigma^2}|\{\widehat{\boldsymbol{\omega}},\widehat{\mathbf{p}}\} = \frac{1}{K}\sum_{k=1}^{K}\frac{\left[y^{\mathrm{M}}(\mathbf{x}^{(k)}) - \mu_{\widehat{y},(-k)}(\mathbf{x}^{(k)})\right]^2}{\sigma^2_{\widehat{y},(-k)}(\mathbf{x}^{(k)})} \tag{34}$$

Where $\sigma^2_{\hat{y},(-k)}(\mathbf{x}^{(k)})$ is the MSE of the GP predictor based on all the training sites except for the $k^{\text{th}}$ set $\mathcal{D}_k$. The most popular version of CV considers sets to include only one sample, i.e. $K = m$. This is called the *Leave-One-Out Cross-Validation* (LOOCV).

**Comparison of MLE and CV**

Martin and Simpson [52] investigated six test problems to compare the performance of MLE and CV for parameter estimation. It was found that MLE was generally better than CV. CV has the potential to perform slightly better, but it also has the risk of performing much worse. In a recent study, Bachoc [Bachoc-2013] also performed numerical study to compare MLE and CV and it was concluded that when the model is mis-specified, CV performs better than MLE. However, MLE is more likely to yield the best predictions as long as the correlation family is properly specified. Therefore, CV is suitable for cases when one gives robustness over best possible performance. For more details and solutions the interested readers are referred to the discussion in [52][68].

**Appendix C. Accuracy Assessment**

The GP metamodel's accuracy needs to be assessed before it can be used in future analyses. The quality of a metamodel is usually assessed using two criteria [52]: (1) accuracy in reproducing the design observations; (2) accuracy of computer model outputs at untried locations. As GP is by construction an interpolator it can reproduce the design samples exactly. Therefore, the first criterion is naturally satisfied. The second criteria can be assessed by measuring the error in GP predictions at independent "validation" or "test" samples. By "independent" we mean that the validation sample set is different from the training sample set. To satisfy this second criterion one may require more simulation runs in addition to those runs used in the raining of metamodels. This can potentially defeat the purpose behind metamodels, which is reducing the computational cost. Through literature review and our own experience, we have observed that it is usually sufficient and most convenient to implement to assess the accuracy using CV and predictivity coefficient. Graphical inspection is also easy to use and provides useful information, details can be found in [55][56].

**Cross Validation and Error Estimation**

A CV procedure can be used to assess the accuracy of metamodels without sampling any additional points beyond those used to train the metamodels. Now we will use $K = m$ (each subset only has one sample) as an example as it is the most popular version of CV. The basic idea is to leave out one observation and predict it back using the metamodel built with the $(m - 1)$ remaining points. By doing this in turn for each training sample we can get the residuals for each prediction, the average of which is called the "Leave-One-Out Cross Validation (LOOCV) error":

$$\mathcal{E}_{\text{LOOCV}} = \frac{1}{m} \sum_{i=1}^{m} \left[ y^M(\mathbf{x}^{(i)}) - \mu_{\hat{y},(-i)}(\mathbf{x}^{(i)}) \right]^2 \tag{35}$$

**Predictivity Coefficient**

The "*Predictivity Coefficient*" $Q_2$ gives the ratio of the output variance explained by the emulator [69] and is given by:

$$Q_2 = 1 - \frac{\sum_{i=1}^{N_{\text{val}}} \left( y^M(\mathbf{x}^{(i)}) - \mu_{\hat{y}}(\mathbf{x}^{(i)}) \right)^2}{\sum_{i=1}^{N_{\text{val}}} \left( y^M(\mathbf{x}^{(i)}) - \overline{y^M} \right)^2} \tag{36}$$

where $N_{\text{val}}$ is the size of the validation sample set, $y^M(\mathbf{x}^{(i)})$ is the output from full model simulation, $\overline{y^M}$ is their empirical mean value and $\mu_{\hat{y}}(\mathbf{x}^{(i)})$ is the predicted output using GP emulator. A $Q_2$ value close to 1.0 means that the GP metamodel is accurate. In practical situations, a metamodel with $Q_2$ value above 0.7 is often considered as a satisfactory approximation of the full model [69]. However, this requires full model executions at a larger sample set than that used to train the GP metamodel.

Equation (36) can be re-written in the following form that incorporates CV errors:

$$Q_2 = 1 - \frac{\sum_{i=1}^{m}\left(y^{\mathrm{M}}(\mathbf{x}^{(i)}) - \mu_{\hat{y},(-i)}(\mathbf{x}^{(i)})\right)^2}{\sum_{i=1}^{m}\left(y^{\mathrm{M}}(\mathbf{x}^{(i)}) - \overline{y^{\mathrm{M}}}\right)^2} \qquad (37)$$

In this case, no further simulator runs are required. Note that $N_{val}$ becomes the size of training sample set $m$. The "cross-validated predictions" $\mu_{\hat{y},(-i)}(\boldsymbol{x}^{(i)})$ are also used to replace $\mu_{\hat{y}}(\boldsymbol{x}^{(i)})$.

**CV vs. generating new test samples**

It is generally up to the user to choose between using CV or generating new test samples for evaluating the accuracy of the GP metamodel. The computational cost should be considered. If a few hundred extra simulator runs are easily affordable, we can just generate new test runs for metamodel validation, this is referred to as "test sample approach" in the literature [68][69]. However, test sample approach can sometimes provide misleading results, especially when the test sample size is small and the test sample locations are not properly chosen. For example, if most of the test samples are generated close to training samples, the GP metamodel will nearly "interpolate" the test samples, giving the wrong impression that the metamodel is accurate. A "sequential validation design" was proposed in [69] that can put test sample points in the unfilled region of the training sample design. This algorithm can optimize the distance between the test set and training set and assess the metamodel predictivity with a minimum number of test samples.

**Appendix D. Implementation Issues**

Many packages or codes have been developed to implement GP metamodeling, with examples including DACE [70], DAKOTA [71], DiceKriging [53], GPML [72], GPM/SA [73][74] and UQLab [54][75]. In the parameter estimation process using MLE, the calculations are preferred to be performed in logarithms to avoid issues with finite precision arithmetic. The correlation matrix **R** needs to be inverted at various stages, e.g., evaluation of the predictor by Equation (14), the regression coefficients by Equation (15), the MSE by Equation (16) or (17). The size of the correlation matrix is $(m \times m)$ which increases with the number of design sites. Inversion of such a matrix is known to suffer from numerical instabilities, especially when two design sites are close, resulting in very similar columns. To solve this problem, a widely used practice is to add a small value $\tau^2$, called *"nugget"* or *"jitter"*, to the diagonal entries of **R**, i.e. $\mathbf{R}_{ii} = 1 + \tau^2$. Such a term serves as a noise factor and it is a convenient way to make sure the covariance matrix is always invertible by introducing negligible errors.

When building the GP emulator, the training inputs are suggested to be scaled between 0 and 1, which correspond to minimum and maximum values from the training set, respectively. Furthermore, the output data can be centralized and standardized so that they have mean 0 and variance 1. Such data processing can reduce the arithmetic errors in matrix inversion and is recommended by [23][24][25][52].

Finally, like almost any other computational tools, an important question about GP metamodeling is how the computational cost increases with the dimensionality of the system? Intuitively higher dimension causes more space between points and the number of training/design points will increase rapidly with the number of inputs. However, in practice computer models never respond strongly to all of their input parameters. Adaptive construction of the GP emulator is possible by progressively assigning more training points to important dimensions which has less smoothness. It was claimed in [44] that GP can be implemented effectively for systems with up to 50 inputs. Considering the booming computational power of the last decade, GP can deal with much higher dimensions on modern computing platforms.

# References

[1] Oberkampf, W. L., & Trucano, T. G. (2002). Verification and validation in computational fluid dynamics. Progress in Aerospace Sciences, 38(3), 209-272.

[2] Oberkampf, W. L., & Roy, C. J. (2010). Verification and validation in scientific computing. Cambridge University Press.

[3] Trucano, T. G., Swiler, L. P., Igusa, T., Oberkampf, W. L., & Pilch, M. (2006). Calibration, validation, and sensitivity analysis: What's what? Reliability Engineering & System Safety, 91(10), 1331-1357.

[4] Cacuci, D. G. (2003). Sensitivity & Uncertainty Analysis, Volume 1: Theory, Chapman and Hall/CRC

[5] Smith, R. C. (2014). Uncertainty quantification: theory, implementation, and applications. SIAM.

[6] Saltelli, A., Ratto, M., Andres, T., Campolongo, F., Cariboni, J., Gatelli, D., & Tarantola, S. (2008). Global sensitivity analysis: the primer. John Wiley & Sons.

[7] Forrester, A. I., & Keane, A. J. (2009). Recent advances in surrogate-based optimization. Progress in Aerospace Sciences, 45(1), 50-79.

[8] Queipo, N. V., Haftka, R. T., Shyy, W., Goel, T., Vaidyanathan, R., & Tucker, P. K. (2005). Surrogate-based analysis and optimization. Progress in Aerospace Sciences, 41(1), 1-28.

[9] Evensen, G. (2009). Data assimilation: the ensemble Kalman filter. Springer Science & Business Media.

[10] Wu, X., & Kozlowski, T. (2017). Inverse uncertainty quantification of reactor simulations under the Bayesian framework using surrogate models constructed by polynomial chaos expansion. Nuclear Engineering and Design, 313, 29-52.

[11] Wu, X., Mui, T., Hu, G., Meidani, H., & Kozlowski, T. (2017). Inverse uncertainty quantification of TRACE physical model parameters using sparse gird stochastic collocation surrogate model. Nuclear Engineering and Design, 319, 185-200.

[12] Unal, C., Williams, B., Hemez, F., Atamturktur, S. H., & McClure, P. (2011). Improved best estimate plus uncertainty methodology, including advanced validation concepts, to license evolving nuclear reactors. Nuclear Engineering and Design, 241(5), 1813-1833.

[13] Campbell, K. (2006). Statistical calibration of computer simulations. Reliability Engineering & System Safety, 91(10), 1358-1363.

[14] Kennedy, M. C., & O'Hagan, A. (2001). Bayesian calibration of computer models. Journal of the Royal Statistical Society: Series B (Statistical Methodology), 63(3), 425-464.

[15] Van Oijen, M., Rougier, J., & Smith, R. (2005). Bayesian calibration of process-based forest models: bridging the gap between models and data. Tree Physiology, 25(7), 915-927.

[16] Gelman, A., Carlin, J. B., Stern, H. S., & Rubin, D. B. (2014). Bayesian Data Analysis (Edition 3). Boca Raton, FL, USA: Chapman & Hall/CRC.

[17] Gilks, W. R., Richardson, S., & Spiegelhalter, D. (1995). Markov chain Monte Carlo in practice. CRC press.

[18] Shrestha, R., & Kozlowski, T. (2016). Inverse uncertainty quantification of input model parameters for thermal-hydraulics simulations using expectation–maximization under Bayesian framework. Journal of Applied Statistics, 43(6), 1011-1026.

[19] Marzouk, Y. M., & Najm, H. N. (2009). Dimensionality reduction and polynomial chaos acceleration of Bayesian inference in inverse problems. Journal of Computational Physics, 228(6), 1862-1902.

[20] Marzouk, Y. M., Najm, H. N., & Rahn, L. A. (2007). Stochastic spectral methods for efficient Bayesian solution of inverse problems. Journal of Computational Physics, 224(2), 560-586.

[21] Ma, X., & Zabaras, N. (2009). An efficient Bayesian inference approach to inverse problems based on an adaptive sparse grid collocation method. Inverse Problems, 25(3), 035013.

[22] Ștefănescu, R., Schmidt, K., Hite, J., Smith, R. C., & Mattingly, J. (2017). Hybrid optimization and Bayesian inference techniques for a non-smooth radiation detection problem. International Journal for Numerical Methods in Engineering.

[23] Arendt, P. D., Apley, D. W., & Chen, W. (2012). Quantification of model uncertainty: Calibration, model discrepancy, and identifiability. Journal of Mechanical Design, 134(10), 100908.

[24] Higdon, D., Kennedy, M., Cavendish, J. C., Cafeo, J. A., & Ryne, R. D. (2004). Combining field data and computer simulations for calibration and prediction. SIAM Journal on Scientific Computing, 26(2), 448-466.

[25] Higdon, D., Gattiker, J., Williams, B., & Rightley, M. (2008). Computer model calibration using high-dimensional output. Journal of the American Statistical Association, 103(482), 570-583.

[26] McFarland, J., Mahadevan, S., Romero, V., & Swiler, L. (2008). Calibration and uncertainty analysis for computer simulations with multivariate output. AIAA journal, 46(5), 1253-1265.

[27] Wilkinson, R. D. (2010). Bayesian calibration of expensive multivariate computer experiments. Large-Scale Inverse Problems and Quantification of Uncertainty, 195-215.

[28] Wilson, G. E. (2013). Historical insights in the development of Best Estimate Plus Uncertainty safety analysis. Annals of Nuclear Energy, 52, 2-9.

[29] Wu, X., Kozlowski, T., Meidani, H., & Shirvan, K., Inverse Uncertainty Quantification using the Modular Bayesian Approach based on Gaussian Process, Part 2: Application to TRACE. (in review)

[30] Han, G., Santner, T. J., & Rawlinson, J. J. (2009). Simultaneous determination of tuning and calibration parameters for computer experiments. Technometrics, 51(4), 464-474.

[31] Bayarri, M. J., Berger, J. O., Paulo, R., Sacks, J., Cafeo, J. A., Cavendish, J., Lin, CH. & Tu, J. (2007). A framework for validation of computer models. Technometrics, 49(2), 138-154.

[32] Brynjarsdóttir, J., & O'Hagan, A. (2014). Learning about physical parameters: The importance of model discrepancy. Inverse Problems, 30(11), 114007.

[33] Liu, F., Bayarri, M. J., & Berger, J. O. (2009). Modularization in Bayesian analysis, with emphasis on analysis of computer models. Bayesian Analysis, 4(1), 119-150.

[34] Ling, Y., Mullins, J., & Mahadevan, S. (2014). Selection of model discrepancy priors in Bayesian calibration. Journal of Computational Physics, 276, 665-680.

[35] Sacks, J., Welch, W. J., Mitchell, T. J., & Wynn, H. P. (1989). Design and analysis of computer experiments. Statistical science, 409-423.

[36] Santner, T. J., Williams, B. J., & Notz, W. I. (2013). The design and analysis of computer experiments. Springer Science & Business Media.

[37] Lancaster, P., & Salkauskas, K. (1981). Surfaces generated by moving least squares methods. Mathematics of computation, 37(155), 141-158.

[38] Buhmann, M. D. (2003). Radial basis functions: theory and implementations (Vol. 12). Cambridge university press.

[39] Schalkoff, R. J. (1997). Artificial neural networks (Vol. 1). New York: McGraw-Hill.

[40] Suykens, J. A., & Vandewalle, J. (1999). Least squares support vector machine classifiers. Neural processing letters, 9(3), 293-300.

[41] Xiu, D., & Karniadakis, G. E. (2002). The Wiener--Askey polynomial chaos for stochastic differential equations. SIAM journal on scientific computing, 24(2), 619-644.

[42] Nobile, F., Tempone, R., & Webster, C. G. (2008). A sparse grid stochastic collocation method for partial differential equations with random input data. SIAM Journal on Numerical Analysis, 46(5), 2309-2345.

[43] Wu, X., Kozlowski, T. & Meidani, H. (2018), Kriging-based Inverse Uncertainty Quantification of Nuclear Fuel Performance Code BISON Fission Gas Release Model using Time Series Measurement Data. Reliability Engineering & System Safety. 169, 422-436

[44] O'Hagan, A. (2006). Bayesian analysis of computer code outputs: a tutorial. Reliability Engineering & System Safety, 91(10), 1290-1300.

[45] Rasmussen, C. E. and Williams, K. K. I (2006). Gaussian processes for machine learning. The MIT Press, 2006. ISBN 0-262-18253-X.

[46] Kleijnen, J. P. (2009). Kriging metamodeling in simulation: A review. European journal of operational research, 192(3), 707-716.

[47] Cressie, N. (2015). Statistics for spatial data, revised edition. John Wiley & Sons.

[48] Matheron, G. (1963). Principles of geostatistics. Economic geology, 58(8), 1246-1266.

[49] Stein, M. L. (2012). Interpolation of spatial data: some theory for Kriging. Springer Science & Business Media.

[50] Currin, C., Mitchell, T., Morris, M., & Ylvisaker, D. (1991). Bayesian prediction of deterministic functions, with applications to the design and analysis of computer experiments. Journal of the American Statistical Association, 86(416), 953-963.

[51] Jones, D. R., Schonlau, M., & Welch, W. J. (1998). Efficient global optimization of expensive black-box functions. Journal of Global optimization, 13(4), 455-492.

[52] Martin, J. D., & Simpson, T. W. (2005). Use of Kriging models to approximate deterministic computer models. AIAA journal, 43(4), 853-863.

[53] Roustant, O., Ginsbourger, D., & Deville, Y. (2012). DiceKriging, DiceOptim: Two R packages for the analysis of computer experiments by Kriging-based metamodelling and optimization. Journal of Statistical Software, 51(1), 54p.

[54] Lataniotis, C., Marelli, S., & Sudret, B. (2017). UQLab User Manual – Kriging (Gaussian Process Modeling). Report UQLab-V1.0-105, Chair of Risk, Safety & Uncertainty Quantification, ETH Zurich.

[55] Wu, X. (2017). Metamodel-based Inverse Uncertainty Quantification of Nuclear Reactor Simulators under the Bayesian Framework, Doctoral dissertation, University of Illinois at Urbana-Champaign.

[56] Bastos, L. S., & O'Hagan, A. (2009). Diagnostics for Gaussian process emulators. Technometrics, 51(4), 425-438.

[57] Chen, W., Xiong, Y., Tsui, K. L., & Wang, S. (2008). A design-driven validation approach using Bayesian prediction models. Journal of Mechanical Design, 130(2), 021101.

[58] Wang, S., Chen, W., & Tsui, K. L. (2009). Bayesian validation of computer models. Technometrics, 51(4), 439-451.

[59] Helton, J. C., Johnson, J. D., Sallaberry, C. J., & Storlie, C. B. (2006). Survey of sampling-based methods for uncertainty and sensitivity analysis. Reliability Engineering & System Safety, 91(10), 1175-1209.

[60] McKay, M. D., Beckman, R. J., & Conover, W. J. (1979). Comparison of three methods for selecting values of input variables in the analysis of output from a computer code. Technometrics, 21(2), 239-245.

[61] Helton, J. C., & Davis, F. J. (2003). Latin hypercube sampling and the propagation of uncertainty in analyses of complex systems. Reliability Engineering & System Safety, 81(1), 23-69.

[62] Johnson, M. E., Moore, L. M., & Ylvisaker, D. (1990). Minimax and maximin distance designs. Journal of statistical planning and inference, 26(2), 131-148.

[63] Morris, M.D. and Mitchell, T.J. 1995 Exploratory designs for computational experiments. Journal of Statistical Planning and Inference 43, 161–174.

[64] Morokoff, W. J., & Caflisch, R. E. (1994). Quasi-random sequences and their discrepancies. SIAM Journal on Scientific Computing, 15(6), 1251-1279.

[65] Niederreiter, H. (1992). Random number generation and quasi-Monte Carlo methods. Society for Industrial and Applied Mathematics.

[66] Bratley, P., Fox, B. L., & Niederreiter, H. (1992). Implementation and tests of low-discrepancy sequences. ACM Transactions on Modeling and Computer Simulation (TOMACS), 2(3), 195-213.

[67] Sobol, I. M. (1976). Uniformly distributed sequences with an additional uniform property. USSR Computational Mathematics and Mathematical Physics, 16(5), 236-242.

[68] Martin, J. D. (2009). Computational improvements to estimating Kriging metamodel parameters. Journal of Mechanical Design, 131(8), 084501.

[69] Iooss B, Boussouf L, Feuillard V, Marrel A. (2010). Numerical studies of the metamodel fitting and validation processes. International Journal of Advances in Systems and Measurements 2010;3:11–21.

[70] Lophaven, S. N., Nielsen, H. B., & Søndergaard, J. (2002). DACE-A Matlab Kriging toolbox, version 2.0.

[71] Adams, B. M. et al (2016). DAKOTA, A Multilevel Parallel Object-Oriented Framework for Design Optimization, Parameter Estimation, Uncertainty Quantification, and Sensitivity Analysis: Version 6.4 User's Manual. Sandia National Laboratories, Tech. Rep. SAND2014-4633.

[72] Rasmussen, C. E., & Nickisch, H. (2010). Gaussian processes for machine learning (GPML) toolbox. Journal of Machine Learning Research, 11(Nov), 3011-3015.

[73] Gattiker, J. (2008), Gaussian Process Models for Simulation Analysis (GPM/SA) Command, Function, and Data Structure Reference, Technical Report LA-UR-08-08057, Los Alamos National Laboratory, Los Alamos, New Mexico.

[74] Williams, B., & Gattiker, J. (2006). Using the Gaussian process model for simulation analysis (GPM/SA) code. Technical report LA-UR-06-5431, Los Alamos National Laboratory, Los Alamos, New Mexico.

[75] Marelli, S. and Sudret B., (2014), UQLab: A framework for uncertainty quantification in Matlab, Proc. 2nd Int. Conf. on Vulnerability, Risk Analysis and Management (ICVRAM2014), Liverpool (United Kingdom), 2554-2563.